\documentclass[default,iicol,sn-nature]{sn-jnl}

\usepackage{graphicx}
\usepackage{adjustbox}
\usepackage{amsmath,amssymb,bm}
\usepackage{siunitx}
\usepackage{braket}
\usepackage{physics}
\usepackage{microtype}
\usepackage{booktabs}
\usepackage{float}
\usepackage[table]{xcolor}
\usepackage{hyperref}
\hypersetup{colorlinks=true,allcolors=blue}

\hypersetup{colorlinks=true,allcolors=blue}

\begin{document}

\title[Ultracoherent superconducting cavity-based multiqudit platform with error-resilient control]{Ultracoherent superconducting cavity-based multiqudit platform with error-resilient control}

\author[1,2]{\fnm{Taeyoon} \sur{Kim}}
\equalcont{These authors contributed equally to this work.}

\author[1]{\fnm{Tanay} \sur{Roy}}
\equalcont{These authors contributed equally to this work.}

\author[1]{\fnm{Xinyuan} \sur{You}}
\author[1]{\fnm{Andy C.~Y.} \sur{Li}}
\author[1]{\fnm{Henry} \sur{Lamm}}
\author[1]{\fnm{Oleg} \sur{Pronitchev}}
\author[1]{\fnm{Mustafa} \sur{Bal}}
\author[1]{\fnm{Sabrina} \sur{Garattoni}}
\author[1]{\fnm{Francesco} \sur{Crisa}}
\author[1]{\fnm{Daniel} \sur{Bafia}}
\author[1]{\fnm{Do\~ga Murat} \sur{K\"urk\c{c}\"uo\~glu}}
\author[1]{\fnm{Roman} \sur{Pilipenko}}
\author[1]{\fnm{Paul} \sur{Heidler}}
\author[1]{\fnm{Nicholas} \sur{Bornman}}
\author[1]{\fnm{David} \sur{van Zanten}}
\author[1]{\fnm{Silvia} \sur{Zorzetti}}
\author[1]{\fnm{Roni} \sur{Harnik}}
\author[1]{\fnm{Akshay} \sur{Murthy}}
\author[1]{\fnm{Andrei} \sur{Lunin}}
\author[3]{\fnm{Sergey} \sur{Belomestnykh}}
\author[1]{\fnm{Shaojiang} \sur{Zhu}}
\author[1]{\fnm{Changqing} \sur{Wang}}
\author[1]{\fnm{Andr\'e} \sur{Valli\`eres}}
\author[1]{\fnm{Ziwen} \sur{Huang}}
\author[2]{\fnm{Jens} \sur{Koch}}
\author[1]{\fnm{Anna} \sur{Grassellino}}
\author*[4]{\fnm{Srivatsan} \sur{Chakram}}\email{schakram@physics.rutgers.edu}
\author*[1]{\fnm{Alexander} \sur{Romanenko}}\email{aroman@fnal.gov}
\author*[1]{\fnm{Yao} \sur{Lu}}\email{yaolu@fnal.gov}

\affil[1]{\orgname{Superconducting Quantum Materials and Systems (SQMS) Center, Fermi National Accelerator Laboratory}, \orgaddress{\city{Batavia}, \state{Illinois}, \postcode{60510}, \country{USA}}}
\affil[2]{\orgname{Department of Physics and Astronomy, Northwestern University}, \orgaddress{\city{Evanston}, \state{Illinois}, \postcode{60208}, \country{USA}}}
\affil[3]{\orgname{Fermi National Accelerator Laboratory}, \orgaddress{\city{Batavia}, \state{Illinois}, \postcode{60510}, \country{USA}}}
\affil[4]{\orgname{Department of Physics and Astronomy, Rutgers University}, \orgaddress{\city{Piscataway}, \state{New Jersey}, \postcode{08854}, \country{USA}}}


\abstract{Realizing the promise of quantum computing requires simultaneously increasing Hilbert-space size and coherence times. While qubit-based architectures have long been the dominant paradigm, large local Hilbert spaces enable qudits, which offer more compact circuits and natural representations for quantum simulation in chemistry~\cite{cao2019quantum}, condensed matter~\cite{Chizzini2024}, and high-energy physics~\cite{Zache:2023cfj}. Superconducting radio-frequency (SRF) cavities are attractive building blocks for qudit-based quantum technologies because of their exceptionally low dissipation and large bosonic Hilbert spaces~\cite{Romanenko2020}, but turning them into programmable modules requires extra effort because the nonlinear circuitry required for control and measurement often introduces loss and noise that erode the memory advantage~\cite{PRXQuantum.5.040307}. Here we demonstrate a two-mode SRF cavity module weakly coupled to an ancillary transmon circuit and engineered to suppress controller-induced dissipation and dephasing, achieving single-photon lifetimes of 20.6\,ms and 15.6\,ms and a dephasing time exceeding 40\,ms. Using sideband interactions together with error-resilient protocols incorporating measurement-based correction and post-selection, we prepare Fock states up to $N=20$ with fidelities exceeding 95\% and generate two-mode entanglement near the coherence limit, approaching 99.9\%. By combining ultracoherent storage with high-fidelity control, this work moves cavity-based hardware beyond memory-only operation and establishes a practical route toward high-dimensional encodings and scalable modular quantum information processing. }


\maketitle
\section{\label{sec:level1}Introduction}

Harmonic oscillators provide a promising complement to qubit-based quantum computing by offering long intrinsic coherence, large local Hilbert spaces, and error channels often dominated by photon loss, enabling hardware-efficient implementations of logical qubits~\cite{mirrahimi2014dynamically,michael2016new}. At the same time, the same bosonic Hilbert space naturally supports qudits, either directly or with error protection~\cite{brock2025quantum}. Qudits increase information density and can reduce circuit depth for many algorithms~\cite{bullock2005asymptotically,wang2020qudits,low2025}. They are particularly well matched to quantum simulation in chemistry~\cite{dutta2024simulating}, condensed matter~\cite{edmunds2024constructing}, and high-energy physics~\cite{gustafson2022primitive,meth2025LG_qudit}, where relevant degrees of freedom such as large spins, rotors, and gauge fields are intrinsically multi-level~\cite{Girvin2024hybrid}. Realizing this opportunity experimentally, however, requires coherent access to high-photon-number manifolds together with control primitives that remain reliable as oscillator coherence improves.

Bosonic quantum systems have advanced rapidly in circuit QED, where coupling to nonlinear superconducting circuits enables universal control and gate operations using a range of techniques~\cite{SNAP2015PRL,Eickbusch2022,rosenblum2018cnot,heeres2017implementing,gao2019entanglement,PRXQuantum.4.020355,LuSchoelkopf2023,Maiti2025} in both single-mode~\cite{reagor2016quantum, Roy2024Qudit} and multimode~\cite{wang2016schrodinger,Chakram2021seamless} platforms. These capabilities have enabled hardware-efficient bosonic quantum error correction, including demonstrations that surpass break-even~\cite{ofek2016extending,sivak2023real,ni2023beating}. In parallel, materials and cavity engineering have pushed oscillator coherence into the tens-of-milliseconds regime~\cite{Rosenblum2023mushroom,Oriani2024Nb_coax}, with further gains expected from improved surface treatments~\cite{Romanenko2020}. This progress sharpens a central challenge: the nonlinear elements that provide control and readout can also introduce loss, dephasing, and backaction, so increasing coherence demands control strategies that remain fast without reintroducing the very errors the cavity is meant to suppress. Approaches such as conditional displacements~\cite{Eickbusch2022}, sideband interactions~\cite{huang2025fast}, and sideband-enhanced SNAP protocols~\cite{you2025floquet} mitigate the speed versus protection tension, but in the high-coherence limit it becomes increasingly important that control be resilient to the dominant ancilla error mechanisms~\cite{pietikainen2024strategies}, especially when targeting high-dimensional manifolds that require many sequential operations.

In this work, we demonstrate a two-mode bosonic module implemented in a two-cell niobium superconducting radio-frequency (SRF) cavity based on the TESLA geometry~\cite{Aune2000, Roy2024Qudit}, weakly coupled to a transmon circuit designed to preserve cavity coherence while enabling efficient control. We achieve single-photon lifetimes of 20.6$\pm$0.4\,ms and 15.6$\pm$0.2\,ms for the two modes, establishing a record for multimode quantum memories. Despite operating in a coherence-preserving weak-coupling regime, we use sideband interactions to implement fast cavity control on timescales far shorter than those set by the dispersive shift. Combined with protocols that make key operations leading-order immune to ancilla errors through measurement-based correction and post-selection, we prepare Fock states up to $N=20$ with fidelities exceeding 95\%, and generate two-mode entanglement with estimated coherence-limited fidelities up to 99.9\% after post-selection. These results provide experimentally validated primitives for high-dimensional bosonic encodings and support a scalable route toward qudit-based quantum computation and simulation in multimode SRF hardware.

\section{Cavity Characterization and System Design}

\begin{figure}
\centering
\includegraphics[width=\columnwidth]{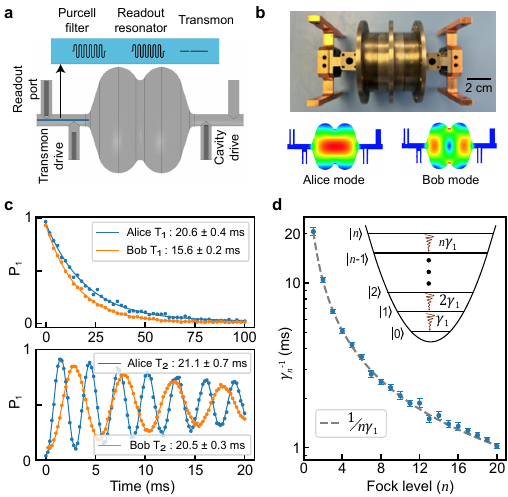}
    \caption{\textbf{Device architecture and coherence properties.} 
    \textbf{a} Schematic of the double-cell elliptical SRF cavity with four ports for RF control. Three ports are used in this experiment. The chip (blue) contains an ancilla transmon, a Purcell filter, and a stripline resonator for measuring the transmon. 
    \textbf{b} Picture of the niobium cavity (top panel) with mounting brackets made of oxygen-free copper. Bottom panel shows the electric field distribution of the two fundamental modes dubbed `Alice' and `Bob'. 
    \textbf{c} Coherence times of the single photon states obtained using T$_1$ and Ramsey experiments. 
    \textbf{d} Relaxation times of Alice mode's Fock levels up to $\ket{20}$ with a fit to $1/(n \gamma_1)$, where $n$ is the Fock state photon number and $\gamma_1=1/T_1^{(1)}$ is the decay rate of $\ket{1}$. The inset shows that the theoretical decay rate of Fock $\ket{n}$ is equal to $n\gamma_1$ for a harmonic oscillator.}
    \label{fig1}
\end{figure}

We designed a two-mode superconducting cavity based on the multi-cell TESLA geometry \cite{Aune2000, Reineri2023}, consisting of two elliptically shaped cells connected by a circular aperture referred to as the iris, as shown in Fig.~\ref{fig1}(a). The nearly degenerate TM$_{010}$ modes of each cell hybridize into symmetric (“Alice”) and antisymmetric (“Bob”) normal modes with similar electric field density at the transmon tunnel, enabling near-equal coupling.
The Alice and Bob modes are designed to have frequencies of 5.779\,GHz and 6.872\,GHz, respectively. This frequency separation helps minimize crosstalk during control operations between the cavity modes, while also providing flexibility in placing the transmon frequency, which can vary due to fabrication inaccuracies. Cylindrical waveguide pipes with sufficiently high cut-off frequencies (35\,GHz) are attached on either side of the cavity; one of these houses the transmon. Additionally, each pipe includes two ports for microwave drive lines to control transmon, readout and cavity modes. 

The cavity is fabricated from high-purity niobium using e-beam welding to eliminate seam loss, followed by chemical and heat treatments to remove surface damage and contaminants. 
The optimized surface quality and the intrinsic bulk property of high-purity niobium enable high quality factors for the cavity modes in the milli-Kelvin and low-photon-number regime~\cite{Romanenko2020, Paul2021, Rosenblum2023mushroom, Roy2024Qudit, Oriani2024Nb_coax}, providing an excellent platform for quantum computing applications.

Using a classical decay measurement, we confirm that the two fundamental modes exhibit exceptionally long decay times of 25.5\,ms and 19.3\,ms, respectively, corresponding to loaded quality factors of approximately $9.4 \times 10^8$ and $8.4\times10^8$.
Temperature-dependent measurements confirm residual resistance and two-level systems (TLS) contributions consistent with SRF best practices (see Supplementary Section II), establishing a high-coherence baseline prior to transmon integration.

To achieve universal control of the two-mode cavity as a quantum processing unit (QPU), we introduce a transmon qubit that simultaneously couples to the two normal modes.
To mitigate inverse Purcell loss, dielectric loss, and seam loss introduced by the transmon, first we engineer the cavity geometry to minimize chip participation, then position the transmon along the cavity axis to reduce dispersive shift while maintaining controllable coupling (see Supplementary Section~III).
 RF simulations confirm that these modifications preserve the TESLA geometry’s inherent low-loss advantage, enabling robust two-mode control via the ancilla qubit without sacrificing the cavity’s long coherence times.

In our experiment, a transmon with frequency $\omega_q/2\pi=$ 6.402\,GHz, relaxation time of $T_1^{ge}=147.4\,\mu$s and Ramsey decay time of $T_2^{ge}=47.3\,\mu$s is integrated with the 2-cell cavity for control and readout. The coupling to the modes results in a dispersive interaction $\left(\chi_e^{a}\hat{a}^\dagger\hat{a} + \chi_e^{b}\hat{b}^\dagger\hat{b}\right)\ket{e}\bra{e}$, where the device parameters are adjusted to obtain similar weak dispersive shifts ($\chi_e^{a,b}/2\pi$) of $–71$\,kHz and $–96$\,kHz for Alice ($\hat{a}$) and Bob ($\hat{b}$). We measure the single-photon lifetime $T_1^{(1)}=1/\gamma_1$ of 20.6\,ms for Alice and 15.6\,ms for Bob, which are reasonably close to the baseline cavity lifetimes. We also measure the $T_2$ of the modes by preparing a $\ket{0}$ and $\ket{1}$ superposition state, finding 21.1\,ms and 20.5\,ms for Alice and Bob modes, respectively. From these measurements, we extract a pure dephasing time of around 43\,ms and 60\,ms for the two modes, which can be accounted for by the transmon's thermal shot noise (see Supplementary Section~I). 

A characteristic property of a linear system is that the decay rate of a Fock state $\gamma_n$ is proportional to its photon number, i.e., $\gamma_n = n \gamma_1$. Equivalently, the lifetime of the $n$-th Fock state satisfies $T_1^{(n)} = T_1^{(1)}/n$. We experimentally verify this relationship by preparing Fock states with increasing photon numbers (as described in the next section) and measuring their relaxation times. Figure~\ref{fig1}(d) shows the measured data in excellent agreement with the theoretical expectation. Remarkably, the lifetime of the $\ket{20}$ state remains above 1\,ms, demonstrating the system’s potential as a high-coherence platform supporting Fock states with large photon numbers, a precursor to qudit control.

\section{Ancilla-Error-Resilient Cavity State Preparation}
\label{sec:state_preparation}
An ancilla qubit dispersively coupled to a linear system enables universal control~\cite{SNAP2015PRA, Girvin2024hybrid}. The resulting photon-number dependence of the ancilla transition frequency enables SNAP gates, which apply arbitrary phases to specific Fock states. Combined with cavity displacements, this allows for universal control of the cavity state~\cite{SNAP2015PRA, SNAP2015PRL}. However, the standard SNAP gate duration should be sufficiently longer than $2 \pi / \lvert\chi_e\rvert$, making it inefficient when $\chi_e$ is small. In contrast, both the Echoed Conditional Displacement (ECD) gate~\cite{Eickbusch2022, You2024crosstalk} and sideband control~\cite{sideband2015PRA, Liu2021qudit, huang2025fast} also enable high-fidelity universal control, with a gate time controlled by microwave drive strength much shorter than the SNAP gate duration limit $2 \pi / \lvert\chi_e\rvert$.

\begin{figure}[hbt!]
    \centering
    \includegraphics[width=\columnwidth]{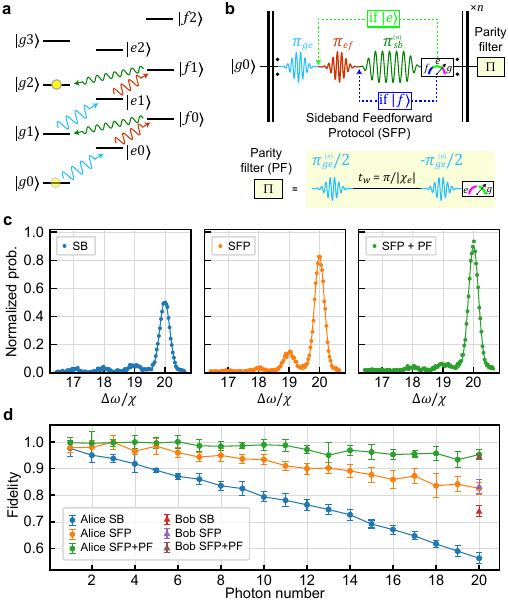}
    \caption{\textbf{Fock state preparation and characterization.} 
    \textbf{a} Sideband scheme demonstrating the transfer of population starting from $\ket{g0}$ to $\ket{gn}$. Addition of one photon in the bosonic ladder involves application of unconditional $\pi_{ge}$ (light blue arrows) and $\pi_{ef}$ (red arrows) pulses on the transmon followed by a $\pi$ pulse ($\pi_{sb}$) which allows the sideband transition $\ket{f,n}\leftrightarrow \ket{g,n+1}$ (green arrows). 
    \textbf{b} Pulse protocol for correcting ancilla errors and cavity single-photon loss. After each $\pi_{sb}$ pulse, the transmon is measured, followed by application of conditional pulses to correct transmon errors, termed as `sideband feedforward protocol' (SFP). At the end of the sequence, a parity filter (PF) is applied to post-select results with the expected parity. The PF is implemented by applying two opposite $\pi_{ge}/2$ pulses conditioned on $n$ photons with a gap of $\pi/|\chi_e|$ so that the correct state is always mapped to $\ket{g}$. 
    \textbf{c} Photon-number-resolved spectroscopy (PNRS) of the $\lvert20\rangle$ state, prepared in the Alice mode, using sideband (SB) only (left), SFP (middle), and both SFP and PF (right). SFP significantly improves the height of the target peak, with remaining infidelity primarily caused by leakage to the off-target state $\lvert19\rangle$. This error is further suppressed by implementing PF and post-selecting on the correct state.   \textbf{d} Fidelities of different Fock states in the Alice (round markers) and Bob (triangular markers for $\ket{20}$ state only) modes prepared using the three methods. The error-resilient methods clearly improve the state preparation fidelities.
    }
    \label{fig2}
\end{figure}

In this experiment, we utilize the sideband (SB) scheme~\cite{sideband2015PRA, huang2025fast} for controlling the cavity and further develop ancilla-error resilient state preparation schemes.
As illustrated in Fig.~\ref{fig2}(a), a large Fock state in a single mode can be prepared by repeatedly applying a sequence of $\pi_{ge}$, $\pi_{ef}$, and $\pi_\mathrm{sb}^{(n)}$ sideband pulses, with each sequence inducing a transition $\ket{g, n} \rightarrow \ket{e, n} \rightarrow \ket{f,n} \rightarrow \ket{g, n+1}$.
The SB-scheme fidelity is primarily limited by the transmon decoherence occurring during sideband pulses since the cavity coherence time significantly exceeds that of the transmon and transmon $\pi$ pulses in our system are approximately twenty times faster than sideband operations.
To mitigate this cumulative infidelity caused by the ancilla, we have developed a ``sideband feedforward protocol" (SFP). Akin to other techniques exploiting multi-level transmon readout for fault-tolerant cavity operations~\cite{Rosenblum2018,Elder2020,Reinhold2020,  teoh2023dual}, the SFP leverages error syndrome detection via high-fidelity single-shot measurements of the transmon populations in the $\ket{g}$, $\ket{e}$, and $\ket{f}$ states after each sideband operation. 
A measurement outcome of $\ket{e}$ ($\ket{f}$) indicates that a decay (dephasing) error has occurred, and a corresponding subsequent corrective pulse is applied as illustrated in Fig.~\ref{fig2}(b).
Furthermore, when climbing the ladder to higher Fock states $N$ using SFP, the qudit relaxation error becomes more significant as its effective rate increases with $N$. Consequently, the remaining off-target population is mostly found in $\ket{N-1}$, as observed experimentally in Fig.~\ref{fig2}(c) using photon-number-resolved spectroscopy (PNRS) and confirmed by an open-system numerical simulation. This makes it possible to apply a parity filter (PF) and filter out by post-selection the $\ket{N-1}$ population for an additional improvement. The details of the SB, SFP, PF, and fidelity estimation protocols are provided in Methods~\ref{meth:sideband_cal_reset_sfp}--\ref{meth:sideband_cal_sfp_eff}, along with the numerical simulation.

\section{Sideband-mediated Virtual Raman Interaction for Cavity Entanglement}

\begin{figure*}[t]
    \centering
    \includegraphics[width=\textwidth]{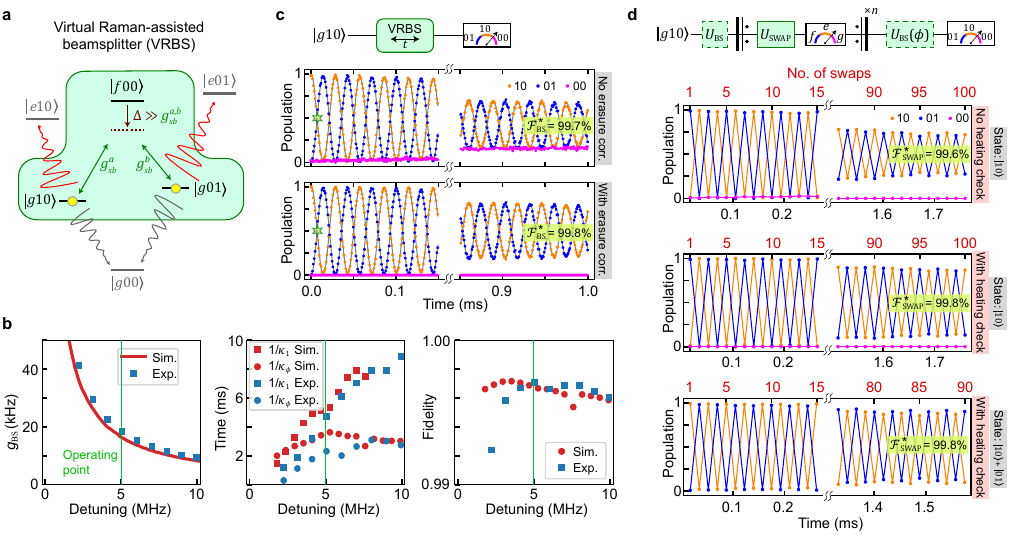}
    \caption{\textbf{Virtual Raman-assisted beamsplitter (VRBS) operation.} 
    \textbf{a} Level diagram showing microwave drives for realizing the VRBS operation between the two modes and relevant stochastic processes that can lead to errors. The green arrows represent detuned $\ket{f0} \leftrightarrow \ket{g1}$ transition for both modes with rates $g_{sb}$. The gray lines show single-photon decay channels giving rise to erasure errors. The red lines represent transmon heating events arising due to sideband drives. 
    \textbf{b} Characterization of VRBS interactions. The three panels show beamsplitter rates, cavity decoherence rates, and gate fidelities as a function of the sideband detuning. Blue and red markers denote experimental data and numerical simulation results, respectively.
    \textbf{c} The time-domain VRBS measurement at the optimal detuning. The top panel shows the circuit diagram, while the middle and bottom panels show data before and after discarding shots containing the erasure error. Fidelities are obtained by fitting to Eq.~\ref{eq:BS_oscillation}.
    \textbf{d} Circuit diagram and the three oscillation graphs showing the VRBS-induced swap operations with heating checks. The swap operations $U_\text{SWAP}$ are applied to the initial state, with a swap time determined from the result in \textbf{c}. The transmon is measured after each swap (top oscillation graph), and a heating check is performed by retaining only shots in which the transmon is found in $\ket{g}$. The gates $U_{\rm BS}$ and $U_{\rm BS}(\phi)$ are only applied for the initialization and readout of the state $(\ket{10}+\ket{01})/\sqrt{2}$. The extracted swap fidelity improves when the heating check is performed, for both the $\ket{10}$ and $(\ket{10}+\ket{01})/\sqrt{2}$ initial states (middle and bottom graph).}
    \label{fig3}
\end{figure*}
As a demonstration of two-mode control in this cavity system, we extend the sideband approach to implement a two-mode entangling operation, realizing a beamsplitter interaction in the single-photon subspace of the two modes. 
Such an interaction has previously been implemented using either a transmon~\cite{Gao2018, pfaff2017controlled} or a dedicated coupling element to enable three- or four-wave mixing between modes~\cite{PRXQuantum.4.020355, LuSchoelkopf2023}. 
In contrast to standard four-wave mixing with a fixed-frequency transmon, we leverage the transmon’s $\ket{f}$ level to mediate the interaction. 
By simultaneously driving the transmon at a detuning $\Delta$ from the $\ket{f0} \leftrightarrow \ket{g1}$ sideband transitions of both Alice and Bob, we activate a virtual-Raman-assisted beamsplitter (VRBS) interaction between the two modes, enabling coherent photon exchange and entanglement (see Fig.~\ref{fig3}(a)). 
Choosing $\Delta$ larger than the sideband rates suppresses direct population of the more lossy transmon $\ket{f}$ state during the interaction, albeit at the cost of a reduced conversion rate. 
Importantly, compared to the standard four-wave mixing protocol, our technique achieves a faster conversion rate for a given drive strength, with an enhancement factor of $|\alpha/\Delta|$, where $\alpha$ is the transmon anharmonicity (see Supplementary Section~IX).

To estimate the fidelity of the entangling operation, we initialize a single-photon Fock state in the Alice mode and monitor the populations of the two-mode states $\ket{10}$, $\ket{01}$, and the erasure state $\ket{00}$ (ordered as Alice $\otimes$ Bob). These populations are measured by mapping cavity states onto distinct levels of the transmon (see Supplementary Section~XIII). The resulting population dynamics are well captured by a fit based on the theoretical model~\cite{LuSchoelkopf2023},
\begin{equation}
P_{\text{Alice}} = \frac{1}{2}e^{-\kappa_1t}(1+e^{-\kappa_{\phi}t} \cos(2g_\mathrm{BS}t)),
\label{eq:BS_oscillation}
\end{equation}
where $g_\mathrm{BS}$ is the beamsplitter rate, with the corresponding beamsplitter time $\pi/(4g_\mathrm{BS})$. 
The effective decay and dephasing rates of the oscillations, $\kappa_1$ and $\kappa_\phi$, together set the upper bounds on the achievable fidelities of the beamsplitter and swap operations, $F^*_\mathrm{BS} \approx 1-\pi\kappa_\mathrm{BS}/(4g_\mathrm{BS})$ and $F^*_\mathrm{SWAP} \approx 1-\pi\kappa_\mathrm{BS}/(2g_\mathrm{BS})$, respectively, where $\kappa_\mathrm{BS}=\kappa_1+\kappa_\phi/2$.
We optimize these fidelities as a function of the VRBS detuning $\Delta$, under fixed sideband drive strengths, $g_{sb}^a/2\pi=269.5\pm0.5$\,kHz, $g_{sb}^b/2\pi=301.8\pm0.5$\,kHz.
The Raman-assisted beamsplitter rate increases as $\Delta$ decreases ($\propto\Delta^{-1}$), as shown by the experimental data in Fig.~\ref{fig3}(b, left). 
However, this enhanced interaction comes at the cost of enhanced decoherence such as the Purcell decay rate ($\propto\Delta^{-2}$). 
On the other hand, for large detuning, the decoherence is dominated by spurious processes from transmon heating (see Discussion), whose rate is largely insensitive to MHz-scale variations in $\Delta$, leading to the observed saturation of the beamsplitter dephasing time $1/\kappa_\phi$ at large detunings in Fig.~\ref{fig3}(b, middle).

The interplay between these mechanisms leads to a maximum in gate fidelity as a function of sideband detuning, as shown in Fig.~\ref{fig3}(b, right). 
We quantitatively capture this trade-off using open-system numerical simulations, from which we extract cavity decoherence times, beamsplitter rates, and the resulting gate fidelities across a range of detunings (see Supplementary Section~X). 
The simulated results show good agreement with experimental measurements.

The coherent photon exchange process corresponding to the VRBS interaction at the optimal parameters is shown in Fig.~\ref{fig3}(c, top). 
Fitting this oscillation using Eq.~\eqref{eq:BS_oscillation} yields a coherence-limited beamsplitter fidelity of $99.686 \pm 0.004 \%$. 
This operation can be interpreted as a single-qubit gate on a dual-rail qubit ($\{\ket{10},\ket{01}\}$) encoded across the single-photon subspace of the two cavity modes~\cite{teoh2023dual, chou2024superconducting}. 
We observe a gradual increase in the population of the $\ket{00}$ state during the interaction as a result of photon loss in the cavity modes, equivalent to the erasure error of a dual-rail qubit. 
While we do not implement an erasure syndrome measurement that preserves the dual-rail qubit state, we estimate the improvement in fidelity from erasure checks by post-selecting on the absence of erasures at the end of the oscillation. 
The resulting oscillation, conditioned on no erasure events, exhibits a prolonged decay constant of $1/\kappa_1 = 70.85$\,ms, translating to an improved $\mathcal{F}^*_\mathrm{BS}$ of $99.810 \pm 0.003 \%$ (Fig.~\ref{fig3}(c, bottom)).

In addition to post-selecting on erasure errors, we also perform mid-circuit detection of transmon heating errors, which, as previously discussed, are the dominant source of dephasing in the VRBS operation.  
These heating errors are detected via a transmon measurement, which has been verified to have negligible backaction on the cavity states due to their weak hybridization (see Supplementary Section~XII).
We demonstrate the resulting fidelity improvement using a sequence of SWAP operations implemented via the same VRBS interaction, which amplifies the effects of heating.
As a baseline, Fig.~\ref{fig3}(d, top) shows the outcome of this sequence\textemdash initialized in $\lvert g10\rangle$\textemdash with only a final erasure check, with measured heating rate approximately 1\% per SWAP (see Supplementary Section~XII).
To quantify the benefit of heating-error detection, we perform transmon measurements after each SWAP; the results (Fig.~\ref{fig3}(d, middle)) show a clear contrast improvement after 100 SWAPs, corresponding to a twofold fidelity enhancement (infidelity reduced from $0.4\%$ to $0.2\%$).
We further verify that this improvement holds for initial states prepared as entangled superpositions of the two modes (i.e., equatorial states on the Bloch sphere in the dual-rail encoding) by applying a $U_{\mathrm{BS}}$ following the initialization of a Fock state in one mode (Fig.~\ref{fig3}(d, bottom)), and another $U_{\mathrm{BS}}(\phi)$ before the readout to account for Stark-shift-induced phase rotations (see Supplementary Section~XI). 

Based on the SWAP gate measurement results, we infer a coherence-limited fidelity approaching $99.9\%$ for the error-detected beamsplitter gate—comparable to state-of-the-art results, albeit with longer gate durations, but achieved using significantly simpler circuit hardware.

\section{Discussion}

This work establishes a two-mode SRF cavity circuit-QED module that combines ultrahigh oscillator coherence with measurement-enabled, error-resilient control. The core message is not simply that SRF cavities offer long lifetimes, but that this coherence can be retained while executing nontrivial, repeatable control primitives. These primitives include high-fidelity access to large Fock manifolds and verified two-mode exchange and entanglement operations, both of which are central ingredients for high-dimensional encodings and modular architectures.

A concrete example is high-photon-number state preparation. In the baseline sideband ladder, decoherence and control errors lead to leakage into lower Fock levels, limiting the peak population of $\ket{20}$ to about 0.51 (Fig.~\ref{fig2}(c)). Incorporating the sideband feedforward protocol (SFP) strongly suppresses the residual population in undesired levels, improving the $\ket{20}$ population to about 0.83 in a representative case (Fig.~\ref{fig2}(c)). This approach relies on robust three-state discrimination of the ancilla, which we achieve at $>98\%$ fidelity using a 1.7~$\mu$s readout pulse (see Supplementary Section~VI). Because the per-step error probability is low, SFP incurs only modest overhead, consistent with the measured efficiency reported in Supplementary Section~VII. When combined with post-selection (PF), the off-target population is further reduced, yielding preparation fidelities of $95.3\pm1.9\%$ and $94.6\pm1.1\%$ for $\ket{20}$ in the Alice and Bob modes, respectively (Fig.~\ref{fig2}(d)). Beyond serving as a benchmark, reliable access to large Fock manifolds is the enabling regime for qudit encodings: sideband control supports arbitrary state synthesis~\cite{hofheinz2009synthesizing} and, in principle, universal multi-qudit operations~\cite{Liu2021qudit}. This opens a route to qudit-based quantum computation~\cite{wang2020qudits,Roy2023two_qutrit,nguyen2024empowering,Litteken2023quantum_waltz}, quantum sensing~\cite{Agrawal2024stimulated,deng2024metrology}, and quantum simulation~\cite{Zache:2023cfj,gustafson2022primitive,meth2025LG_qudit} in a hardware setting where long oscillator coherence is a first-class resource.

We further demonstrate a virtual-Raman beamsplitter (VRBS) interaction mediated by the intermediate $\ket{f00}$ state, enabling coherent photon exchange and entanglement within the single-photon subspace, with coherence-limited beamsplitter fidelities up to 99.9\% after post-selection. While the present gate speed is limited by the achievable sideband rates, this is an engineering constraint rather than a fundamental limitation, and it can be improved through device design and wiring optimizations. At the same time, the VRBS measurements clarify the dominant error mechanism at large detunings: cavity decoherence becomes dominated by dephasing associated with drive-induced transmon heating. Such heating can arise from undesired parametric processes such as dressed dephasing~\cite{Boissonneault2008}. A heating event excites the transmon, effectively interrupting coherent exchange, and random heating leads to dephasing because the beamsplitter rate depends strongly on the ancilla state (see Supplementary Section~X). Heating also enhances photon shot-noise dephasing in the cavities. Importantly, this error channel is detectable. By inserting mid-circuit ancilla measurements, we realize a partially error-detected SWAP operation that can be used to suppress dominant dephasing events in shallow-depth algorithms and to support dual-rail surface-code implementations using simpler fixed-frequency transmon couplers. Related calibration strategies can extend this error-detection capability to beamsplitter operations by compensating differential Stark-shift phases~\cite{LuSchoelkopf2023}.

Extending beyond the single-photon manifold, the VRBS interaction continues to couple states within fixed total photon-number sectors and mediates single-photon exchange between modes. The transmon nonlinearity then introduces photon-number-dependent corrections to the beamsplitter rate because the effective detuning is modified by dispersive shifts in higher-photon-number sectors (see Supplementary Section~IX). This yields a nonlinear (non-Gaussian) beamsplitter that can serve as a primitive for qudit entangling gates and for simulations of interacting bosonic systems. As one example, in the two-qudecit ($d=10$) subspace, thirty VRBS operations combined with single-qudecit rotations can synthesize a qudecit CSUM gate with 96\% fidelity (Supplementary Section~XIV). More broadly, these observations motivate extending feedforward beyond state preparation to gate operations, where detected ancilla events can be used to reject or correct trajectories that would otherwise dominate infidelity.

Finally, the two-cell elliptical architecture is naturally extensible along two complementary scaling directions. Increasing the number of TESLA cells yields larger multimode systems that can be configured as cascaded random-access quantum memories~\cite{li2025cascaded}, with additional gains expected from continued SRF surface-treatment advances~\cite{Romanenko2020,Posen2020}. In parallel, two-cell modules can serve as networkable building blocks for modular quantum systems, using quantum-state routing~\cite{zhou2022modular} and tunable couplers~\cite{LuSchoelkopf2023,PRXQuantum.4.020355,Maiti2025} for exchange of quantum information and distribution of entanglement. Together, these scaling paths point to architectures that combine long-lived multimode storage with verified control primitives, enabling high-dimensional quantum processing and sensing at circuit-relevant timescales.


\bibliography{main}

\section{Methods}

\numberwithin{equation}{section}
\setcounter{equation}{0}

\subsection{Cavity fabrication, loss characterization, and mitigation of transmon-induced losses}
\label{meth:fabrication_loss_transmon}

The two-cell niobium cavity is CNC-machined from high-purity material (RRR $\sim 300$) in three parts and electron-beam welded along the equators to avoid seams at the joints. After buffered chemical polishing (BCP) removing $\sim120~\mu$m, the cavity is high-pressure water rinsed and vacuum-baked at $800^\circ\mathrm{C}$ for 3~h to degas hydrogen and reduce processing-related loss.

We characterize bare-cavity relaxation and dephasing in the milli-Kelvin regime using ringdown measurements (200~ms drive followed by free decay) measured in reflection with a vector network analyzer (see SI Section II). A phase-insensitive average yields the energy relaxation time,
\begin{align}
P_\text{PI} \propto\langle I^2 + Q^2 \rangle &\propto e^{-\frac{t}{T_1}}.
\label{eq:phase_insensitive_avg_meth}
\end{align}
A phase-sensitive average provides an effective decay that includes dephasing,
\begin{align}
P_\text{PS}\propto\langle I \rangle^2 + \langle Q \rangle^2 
&\propto e^{-\frac{t}{T_2/2}},
\label{eq:phase_sensitive_avg_meth}
\end{align}
where the approximation assumes frequency-independent phase-noise spectral density over the relevant bandwidth. The corresponding relation is
\begin{align}
\frac{T_2}{2} = \left(T_1^{-1}+2T_\phi^{-1}\right)^{-1},
\label{eq:T2_ringdown_meth}
\end{align}
with $T_\phi$ the intrinsic dephasing time. 

To separate intrinsic cavity losses from temperature-dependent mechanisms, we measure the internal quality factor as a function of temperature and fit it to a standard two-level-system (TLS) plus residual-resistance model,
\begin{align}
\frac{1}{Q_0(T)}=F\delta_0 \tanh\Big(\beta \frac{\hbar \omega}{2k_BT}\Big)+\frac{R_{\text{res}}}{G}.
\label{eq:Q_vs_T_meth}
\end{align}
Here, $F$ is the surface oxide filling factor and $G$ is the geometry factor, computed from finite-element simulations via
\begin{align}
F = \frac{\int_{\text{ox}} \epsilon_{\text{ox}}|\vec{E}|^2dv}{\int_{\text{all}} \epsilon_{\text{0}}|\vec{E}|^2dv},
\label{eq:F_meth}
\end{align}
and
\begin{align}
G=\omega \mu_0 \frac{\int_{\text{all}}|\vec{H}|^2dv}{\int_{\text{sur}}|\vec{H}|^2d\sigma}.
\label{eq:G_meth}
\end{align}
The fitted parameters (see Supplementary Section~II) guide mitigation strategies, including additional BCP to remove embedded contaminants and improved surface treatments to suppress TLS loss.

To preserve cavity coherence after integrating the transmon controller, the transmon chip is inserted through a narrow, small-diameter end tunnel that forms a high-cutoff cylindrical waveguide, enabling selective coupling while suppressing additional radiation and seam-loss channels. Fast readout is enabled by an on-chip Purcell filter, allowing a readout resonator linewidth $\kappa_r \approx 0.5$~MHz while keeping readout-limited linewidths small for the qubit ($\kappa_q \approx 20$~Hz) and cavity modes ($\kappa_{a,b} \approx 0.02$~Hz).

We estimate transmon-chip-induced cavity losses using finite-element simulations of participation ratios for bulk dielectric and relevant surface regions. With surface thickness $t_{\mathrm{surf}}=3$~nm and relative permittivity $\epsilon_r=10$, the participation ratios are
\begin{align}
p_{\text{bulk}} &= \frac{\int_{\text{bulk}}\epsilon|\vec{E}|^2dv}{\int_{\text{all}}\epsilon|\vec{E}|^2dv},
\label{eq:pbulk_meth}
\end{align}
\begin{align}
p_{\text{MA}}&=\frac{t_{\text{surf}}\int_{\text{MA}}\epsilon_0|\vec{E}|^2d\sigma}{\epsilon_{\text{r,MA}}\int_{\text{all}}\epsilon|\vec{E}|^2dv},
\label{eq:pMA_meth}
\end{align}
and
\begin{align}
p_{\text{SA,MS}}&=\frac{t_\text{surf}\int_{\text{SA,MS}}\epsilon|\vec{E}|^2d\sigma}{\int_{\text{all}}\epsilon|\vec{E}|^2dv}.
\label{eq:pSAMS_meth}
\end{align}
Using these participation ratios and loss tangents, we estimate $\kappa_{\text{bulk}} = \omega_c p_{\text{bulk}}\tan\delta_{\text{bulk}}$ and $\kappa_{\text{surf}}=\omega_c\sum_{j=\text{MS,MA,SA}}p_j\tan\delta_j$. We target dispersive shifts for each cavity mode in the 50--100~kHz range to balance fast sideband control against inherited loss; under our operating parameters, transmon-material losses are not the dominant limitation of cavity $T_1$.

When the Lamb shift is comparable to the bare transmon--cavity detuning, accurate loss estimates require accounting for nonlinearity-induced frequency renormalization. We therefore use an energy-participation-ratio (EPR) analysis to obtain the Lamb-shifted transmon frequency and renormalize the Josephson inductance used in the participation-ratio simulations, enabling a consistent estimate of inherited loss.

\subsection{SFP protocol}
\label{meth:sideband_cal_reset_sfp}
For the SB scheme, the transmon-resonator system is initially in state $\ket{g,0}$. The transmon is unconditionally excited to the $\ket{f}$ level by applying broadband $\pi_{ge}$ and $\pi_{ef}$ pulses. Then a sideband $\pi_\mathrm{sb}^{(0)}$ pulse at the $\ket{f,0} \leftrightarrow \ket{g,1}$ transition is applied to prepare the state $\ket{g1}$ (see Supplementary Section~IV for the calibration of the sideband pulse). By repeatedly applying a sequence of $\pi_{ge}$, $\pi_{ef}$, and $\pi_\mathrm{sb}^{(n)}$ sideband pulses at $\ket{f,n} \leftrightarrow \ket{g,n+1}$, it is possible to prepare Fock states up to a large photon number. It's worth noting that the sideband pulses need not be frequency selective for Fock state preparation, or even for certain superposition states utilizing a recently developed shelving technique~\cite{huang2025fast}. The sideband operation also allows for driven-dissipative reset of the cavity state in a time scale much faster than its natural lifetime (see Supplementary Section~V), which is utilized throughout our experiment. 

The dominant source of error during Fock preparation using the SB scheme is the transmon decoherence. Specifically, a transmon decay event from the initial state $\ket{f,n}$ results in the error state $\ket{e,n}$, preventing the sideband transition to $\ket{g,n+1}$; similarly, transmon dephasing disrupts the coherent transition $\ket{f,n}$ to $\ket{g,n+1}$, leaving residual population in $\ket{f,n}$. Although the probability of such events within a single sideband operation is low, the multiple repetitions required to prepare high-photon-number Fock states can cumulatively lead to significant infidelity.

The SFP's error syndrome detection result is fed forward to determine the subsequent corrective pulse in real time. A measurement outcome of $\ket{f}$ heralds a dephasing event, which we correct by repeating the same sideband pulse. An outcome of $\ket{e}$ indicates a transmon decay event; in this case, we first apply a $\pi_{ef}$ pulse and then repeat the sideband pulse to restore the intended state. Finally, an outcome of $\ket{g}$ indicates the ideal scenario where no error has occurred, and thus requires no further correction.

\subsection{Parity filter}
\label{meth:sideband_cal_parity_filter}
For the PF, we apply a $\pi_{ge}/2$ pulse at the frequency $\omega_0 + N \chi$, wait for $\pi/|\chi_e|$, then apply a $(-\pi_{ge})/2$ pulse, and finally perform a transmon readout. During the wait time of $\pi/|\chi_e|$, the transmon acquires a phase shift dependent on whether the total photon number in the cavity is even or odd. The final $(-\pi_{ge})/2$ pulse maps correct parity states to $\ket{g}$ and incorrect parity states to $\ket{e}$. We then discard measurement outcomes where the transmon is found in $\ket{e}$, indicating a parity error.

\subsection{Fock state fidelity estimation}
\label{meth:sideband_cal_fidelity_estimation}
In order to estimate the state preparation fidelity $\mathcal{F}_n$ of a Fock state with photon number $n$, one can avoid full tomography as $\mathcal{F}_n$ is equal to the population of $\ket{n}$. We measure the cavity Fock state occupation using photon-number-resolved spectroscopy (PNRS), where the probabilities are normalized using the transmon's readout ($\ket{g}$ and $\ket{e}$) values.

We perform an open-system simulation for the Fock state preparation protocol, and details are provided in the Supplementary Information. The simulation shows that SFP effectively corrects the transmon decoherence errors, which are the main source of infidelities in the sideband (SB) case. While the SFP fidelity is noticeably improved by correcting the transmon decoherence errors, the remaining infidelity arises primarily from transmon readout errors and qudit relaxation errors caused by the additional readout operations. This change in the source of infidelities explains the out-of-target population predominantly observed in the state $|N-1\rangle$ when using SFP compared to a more uniform out-of-target distribution in the SB case, as shown in Fig.~\ref{fig2}(c), since the effective qudit relaxation rate increases with the photon number while the transmon decoherence errors are largely insensitive to the photon number. The simulation results indicate that the fidelities of SFP can be further enhanced by optimizing transmon readout and reducing qudit decoherence, particularly when preparing high-photon-number states.

\subsection{Sideband calibration \& SFP efficiency}
\label{meth:sideband_cal_sfp_eff}
The cavity--transmon sideband drive is implemented as a flat-top pulse with 40~ns squared-sine ramps on each edge to improve adiabaticity. Because the sideband transition experiences a drive-amplitude-dependent Stark shift, the operating transition frequency can deviate from the bare resonance $2\omega_q + \alpha-\omega_c$. We therefore calibrate the sideband frequency by sweeping the drive frequency around the nominal resonance and monitoring the resulting $\ket{f}$-state population; the resonance is identified by a pronounced drop in $\ket{f}$ population. After locating the resonance, we calibrate the sideband $\pi$-pulse duration by measuring time-domain Rabi oscillations on the $\ket{f,n}\leftrightarrow\ket{g,n+1}$ transition. This calibration can be performed together with the SFP and PF protocols to retain contrast at high photon number.

We quantify the overhead of the sideband-feedforward protocol (SFP) by measuring how often additional feedforward operations are needed per step when preparing a target Fock state. Averaged over many trials, we apply feedforward $1.059\pm0.003$ times per ladder step, indicating minimal time overhead. We observe that 0.49\% of shots cannot be corrected by SFP, consistent with errors outside the correctable subspace (for example, photon loss events in the cavity). These measurements confirm that SFP provides a substantial fidelity gain at low overhead under our operating conditions.

\subsection{Sideband-assisted cavity reset}
We also implement a fast cavity reset based on engineered dissipation mediated by the sideband interaction. The sideband drive hybridizes cavity excitations with the transmon, so that the cavity population inherits the transmon’s much faster decay. In the regime where the sideband interaction strength exceeds the transmon decay rate, the effective decay from $\ket{g,n}$ proceeds through the cascaded process $\ket{g,n}\rightarrow\ket{f,n-1}\rightarrow\ket{e,n-1}\rightarrow\ket{g,n-1}$ at a rate set by the transmon relaxation (up to an $\mathcal{O}(1)$ factor)~\cite{Lu2017}. To reset an arbitrary initial cavity state, we apply a sequence of sideband drives addressing photon numbers from $\ket{g,N_{\max}}\leftrightarrow\ket{f,N_{\max}-1}$ down to $\ket{g,1}\leftrightarrow\ket{f,0}$, with each tone applied for a duration long compared to the relevant transmon decay time. This autonomous protocol does not require knowledge of the exact initial cavity state and is tolerant to small sideband detunings. For a representative coherent state in the Alice mode (initialized with $\alpha=4$), choosing $N_{\max}=26$ ensures the cumulative population above $N_{\max}$ is below 0.5\%. Using a 1~ms duration per sideband tone, the protocol reduces the mean photon number below 0.005 in 25~ms, compared to $\sim$160~ms under natural decay. The reset can be further accelerated by increasing the engineered damping, for example via parametric coupling to a strongly damped readout mode.

\subsection{VRBS calibration}
\label{meth:vrbs_cal_heat_map}

Accurate VRBS operation requires satisfying the frequency-matching condition in the presence of drive-induced Stark shifts, so the effective VRBS detuning generally differs from the bare Alice--Bob detuning. We calibrate the matching condition by preparing the Alice mode in $\ket{1}$ and simultaneously applying the $\ket{f00}\leftrightarrow\ket{g10}$ and $\ket{f00}\leftrightarrow\ket{g01}$ sideband drives with the same detuning at the operating amplitudes. We keep the Alice sideband drive frequency fixed while sweeping the Bob sideband drive frequency and identify the resonance condition by maximizing the measured Bob $\ket{1}$ population.

A small difference in the Stark shifts of the two drives produces a relative phase rotation between Alice and Bob that can accumulate during idle periods between concatenated VRBS pulses and during ramps~\cite{LuSchoelkopf2023}. In addition, for error-detected VRBS sequences, repeated ancilla measurements can contribute to a differential phase rotation because Alice and Bob have different dispersive couplings to the readout mode. If uncompensated, these phases become coherent errors that reduce the fidelity of dual-rail operations such as swaps within the single-photon manifold. We measure the net phase accumulated over repeated VRBS swap operations by initializing the system in $(\lvert 01\rangle+\lvert 10\rangle)/\sqrt{2}$ and varying the phase $\phi$ of the final beamsplitter operation that maps the state back to a single-photon excitation in Alice; the optimal $\phi$ yields the accumulated phase. We observe a linear increase of the accumulated phase with the number of swap operations and extract a phase shift per gate of $0.0046$~rad, which is compensated by applying the corresponding phase in the beamsplitter operation for error-detected swaps.

\subsection{Mid-circuit transmon heating detection}
Transmon heating events during VRBS swaps interrupt coherent oscillation between Alice and Bob and lead to dephasing through stochastic hopping of the ancilla state, which changes the effective VRBS rate due to its dependence on the ancilla state. We therefore perform a transmon measurement after each swap operation to detect heating events. By repeating sequences of up to 100 consecutive swaps and post-selecting on no heating, we fit the cumulative detection probability to $P(n)=1-(1-P_\uparrow)^n$ to extract the heating probability per swap, obtaining $P_\uparrow \approx 1.167\pm0.003\%$ per swap. To verify that repeated ancilla readouts do not measurably degrade cavity lifetime, we prepare the cavity in $\ket{1}$, apply $N$ consecutive transmon readouts (each of duration $t_r=4~\mu$s separated by $\Delta t=3.4~\mu$s), and then map the remaining cavity $\ket{1}$ population onto the transmon for measurement. Converting to elapsed time $t=N(t_r+\Delta t)$, we extract $P_1(t)$ and fit to an exponential; within experimental uncertainty, the resulting decay time agrees with the cavity $T_1$ measured without interleaved readouts.

\subsection{Dual-rail state mapping}
To perform single-shot readout of dual-rail single-photon populations in Alice and Bob, we map the cavity states onto transmon states. After resetting the transmon to $\ket{g}$, we apply a sideband pulse on $\ket{g10}\leftrightarrow\ket{f00}$ (Alice), followed by a photon-number-unselective $\pi_{ef}$ pulse with short duration ($\Delta t \ll 2\pi/\chi_e^a,\,2\pi/\chi_e^b$), and then a second sideband pulse on $\ket{g01}\leftrightarrow\ket{f00}$ (Bob). This sequence maps $\ket{10}\mapsto\ket{e}$, $\ket{01}\mapsto\ket{f}$, and $\ket{00}\mapsto\ket{g}$, enabling single-shot discrimination of $\ket{00}$, $\ket{10}$, and $\ket{01}$ on timescales $\ll 1/\chi_e$. The resulting confusion matrix for this mapping is
\begin{equation}
\label{eq:map_confusion}
M_{\rm map} = 
\begin{bmatrix}
0.9959 & 0.0491 & 0.0100\\
0.0018 & 0.9467 & 0.0172\\
0.0022 & 0.0042 & 0.9728
\end{bmatrix},
\end{equation}
where columns (rows) correspond to prepared (measured) states ordered as $\ket{00}$, $\ket{10}$, and $\ket{01}$. The observed asymmetry between Alice and Bob arises from the order of the mapping sequence. We use $M_{\rm map}$ to correct readout errors in reported dual-rail populations for beamsplitter operations.

\subsection{Qudit gates}
\label{app:qudit_gates}

Universal control of a two-qudit system requires the ability to perform arbitrary single-qudit rotations along with at least one entangling operation~\cite{Braunstein2005CV}. In our multimode architecture, single-qudit rotations $R_d(\vec{\theta})$ can be implemented using either sideband gates~\cite{Liu2021qudit} or echoed conditional displacement gates~\cite{You2024crosstalk}. The VRBS interaction discussed in the main text provides a mechanism for generating entanglement between higher Fock states as well. This operation takes a block-diagonal form, where each basis state $\ket{m,n}$ is rotated within a photon-number-conserving subspace—i.e., among states $\ket{m',n'}$ satisfying $m'+n' = m+n$. Together with single-qudit rotations, this operation enables universal control within a finite-dimensional Hilbert space constructed from $d$-dimensional qudits.  For example, universal operations in the two-qudit Hilbert space can be performed by:
\begin{align}
\label{eq:universal_qutrit}
    \mathcal{U}= & \prod_{j=1}^{N_{\rm block}}\left[\textsc{R}_d(\vec{\theta}_j) \otimes \textsc{R}_d(\vec{\phi}_j) \cdot \textsc{VRBS} \right] \nonumber \\
    &\ \cdot \textsc{R}_d(\vec{\theta}_0) \otimes \textsc{R}_d(\vec{\phi}_0),
\end{align}
where $N_{\rm block}$ is the number of blocks of those gates within the square brackets. 

Using the simulated results discussed in SI Section IX (Fig. S12), it is possible to investigate the properties of VRBS as a two-qudit entangling gate with dimension $d=10$, as a concrete example.
For $d=10$ two-qudit space, the VRBS gate has operator Schmidt rank $100$. The entangling power, $e_p(U)$ can be defined as the average of the linear entropy over product states $\ket{\Psi}=\ket{\psi}_1\otimes\ket{\psi_2}$ following~\cite{wang2003entangling}:
\begin{equation}
    E(\ket{\Psi})=1-\text{Tr}_1 \rho_1^2,
\end{equation}
where $\rho_1=\text{Tr}_2(U\ket{\Psi}\bra{\Psi}U^\dagger)$ is the reduced density matrix, and $U$ is the entangling gate being investigated. The maximum entangling power in $d$ dimensions is $e_p^{max}=\frac{d^2 - d}{d^2 + 1}$, which for $d=10$ is 0.891. We estimate $e_p(\rm VRBS)=0.780(1)$ by sampling $10^3$ $\ket{\Psi}$, which is larger than the qudecit CSUM gate (${\rm CSUM_{10}}|m,n\rangle = \ket{m,m+n \ {\rm mod} \ 10 }$) which is known analytically to have $e_p(\text{CSUM}_{10})=90/121\sim 0.743$~\cite{wang2003entangling}. This implies that VRBS is an efficient qudecit entangling gate.  Further, by numerical optimization of the $SU(10)$ angles in Eq.~(\ref{eq:universal_qutrit}), approximate synthesis of the CSUM$_{10}$ gate from $N_{\rm block}=[10,16,26,30]$ blocks were identified with fidelities
\begin{equation}
    \mathcal{F}_{\text{CSUM}_{10}}^{N_{\rm blocks}}=\{64\%,79\%,93\%,96\%\}.
\end{equation}
Demonstration of universality in our 2-cell system is the topic of future work, with emphasis placed on qudit CSUM gates.


%

\section{Data availability}
Data supporting the findings of this study are available from the corresponding authors upon reasonable request.

\section{Code availability}
Custom code used in this study is available at \href{https://dx.doi.org/10.5281/zenodo.20838677}{10.5281/zenodo.20838677}.

\section{Acknowledgments}
The authors acknowledge Edward D. Pieszchala, Tim Ring, Davida Smith, Damon J. Bice, Michael H. Foley, Ryan Treece, Charles J. Grimm, Scott D. Adams, Theodore Ill, Mackenzie Ring, and Dominic Baumgart for their support in cavity fabrication, processing, and assembly; David I. Schuster for providing the JPA for the measurement; Kevin A. Villegas Rosales, Michal Goldenshtein, Taekwan Yoon, and Vikrant Mahajan for their valuable assistance with instrument setup and coding for measurements; and Zachary Goff-Eldredge and John W. O. Garmon for insightful discussions. The work was supported by the U.S. Department of Energy, Office of Science, National Quantum Information Science Research Centers, Superconducting Quantum Materials and Systems (SQMS) Center under the contract No. DE-AC02-07CH11359. This work made use of the Pritzker Nanofabrication Facility, part of the Pritzker School of Molecular Engineering at the University of Chicago, which receives support from Soft and Hybrid Nanotechnology Experimental (SHyNE) Resource (NSF ECCS-2025633), a node of the National Science Foundation’s National Nanotechnology Coordinated Infrastructure {[RRID: SCR\_022955]}.

\section{Author contributions}
Y.L. conceived and led the project, designed the experimental architecture, and developed the key control protocols. T.K. carried out the core device design and simulations, performed the primary measurements, and analyzed data. T.R. contributed to experimental protocols, JPA design and control, data analysis, and manuscript preparation. X.Y. and A.C.Y.L. developed theoretical and numerical analyses for the VRBS operations and the feedforward protocol. H.L. contributed qudit theory and CSUM gate analysis. O.P. contributed mechanical design and fabrication support. M.B., S.G., and F.C. supported the fabrication of ancilla devices. D.B. contributed to the materials loss analysis. D.M.K. contributed to qudit-related writing and discussion. R.P., P.H., N.B., and D.v.Z. supported RF measurement electronics. S.C. contributed to experimental protocols for cavity control and measurements, as well as shielding, filtering, and thermal-noise mitigation. A.R., Y.L., S.C., J.K., T.R. and A.G. supervised the project. All authors contributed to discussions and manuscript preparation.

\section{Competing interests}
The authors declare no competing interests.

\end{document}


\title{Supplementary Information for ``Ultracoherent superconducting cavity-based multiqudit platform with error-resilient control"}

\maketitle
\beginsupplement

\section{System Hamiltonian parameters}
\label{app:system_parameters}
In this experiment, the relevant bosonic modes include two 3D high-Q cavity modes (Alice and Bob), one quasi-planar readout resonator mode, and the weakly nonlinear ancilla transmon mode. The system Hamiltonian can be expressed as 
\begin{equation}
\begin{split}
\hat{\mathcal{H}}/{\hbar} =\ &
\omega_a \hat{a}^\dagger \hat{a} 
+ \omega_b \hat{b}^\dagger \hat{b}
+ \omega_r \hat{r}^\dagger \hat{r}\\
&+ \omega_q \ket{e}\bra{e} + (2\omega_q+\alpha)\ket{f}\bra{f}\\
& + \chi_e^a\, \ket{e}\bra{e}\hat{a}^\dagger \hat{a}
+ \chi_{e}^b\, \ket{e}\bra{e}\hat{b}^\dagger \hat{b} \\
&+ \chi_{e}^r\, \ket{e}\bra{e}\hat{r}^\dagger \hat{r}.
\end{split}
\end{equation}
Values of the system parameters are shown in Table~\ref{table:calibration_data}. 
The inverse-Purcell limit on the cavity mode lifetime is determined by both the relaxation as well as the dephasing channel of the transmon, which can be approximated as~\cite{Rosenblum2023mushroom}
\begin{align}
    T^\mathrm{Purcell}_{1a,b} \approx \left(\frac{\Delta_{a,b}}{g_{a,b}}\right)^2\frac{T_{2E}^{ge}}{2}.\label{eq:Purcell}
\end{align}
From this, we estimate decay times of $T^\mathrm{Purcell}_{1a} = 1.17\,$s and $T^\mathrm{Purcell}_{1b}=0.34\,$s, far exceeding the bare Alice and Bob lifetimes limited by the cavity's intrinsic material loss. It's worth noting that this approximation assumes the same dephasing rate measured from the Hahn-echo experiment for the actual dressed-dephasing-induced cavity decay process. In reality, such a decay process is sensitive to dephasing noise at the detuning between the transmon and the cavity, which can have very different spectral components than in the Hahn-echo experiment. A more precise understanding of the inverse-Purcell limit due to the dephasing noise would require probing the dephasing spectral density via spin-locking techniques~\cite{Yan2013}.

We also estimate the cavity dephasing time limit due to the transmon thermal shot noise. In the small thermal population limit, we approximate the thermal-noise-induced cavity dephasing time as~\cite{Clerk2007}
\begin{equation}
    T^\mathrm{th}_{\phi a,b} = \frac{(\chi_e^{a,b})^2+(T_1^{ge})^{-2}}{\bar{n}^q_\mathrm{th}(T_1^{ge})^{-1}(\chi_e^{a,b})^2}.
\end{equation}
Plugging in parameters in Table~\ref{table:calibration_data}, we obtain $T^\mathrm{th}_{\phi} \approx 59$\,ms for both modes, reasonably close to the measured cavity dephasing times.

\begin{table}[t]
\centering
\small 
\rowcolors{2}{gray!25}{white}
\begin{adjustbox}{max width=\columnwidth}
\begin{tabular}{c|c|c}
\toprule
\rowcolor{gray!50}
\textbf{Parameter} & \textbf{Description} & \textbf{Value} \\
\midrule
$\omega_a$ & Alice mode frequency & $2\pi \times 5.779\,\mathrm{GHz}$ \\
$T_1^A$ & Alice mode relaxation time & $20.6\,\mathrm{ms}$ \\
$T_2^A$ & Alice mode Ramsey decoherence time & $21.1\,\mathrm{ms}$ \\
$T_\phi^A$ & Alice mode pure dephasing time & $43.2\,\mathrm{ms}$ \\
$\bar{n}_{\text{th}}^a$ & Alice mode thermal population & $< 0.1\%$ \\
$\omega_b$ & Bob mode frequency & $2\pi \times 6.872\,\mathrm{GHz}$ \\
$T_1^B$ & Bob mode relaxation time & $15.6\,\mathrm{ms}$ \\
$T_2^B$ & Bob mode Ramsey decoherence time & $20.5\,\mathrm{ms}$ \\
$T_\phi^B$ & Bob mode pure dephasing time & $59.8\,\mathrm{ms}$ \\
$\bar{n}_{\text{th}}^b$ & Bob mode thermal population & $< 0.1\%$ \\
$\omega_q$ & Transmon $|g\rangle \leftrightarrow |e\rangle$ frequency & $2\pi \times 6.402\,\mathrm{GHz}$ \\
$\alpha$ & Transmon anharmonicity & $-2\pi \times 245\,\mathrm{MHz}$ \\
$T_1^{ge}$ & Transmon decay time & $147.4\,\mu\mathrm{s}$ \\
$T_2^{ge}$ & Transmon Ramsey dephasing time & $47.3\,\mu\mathrm{s}$ \\
$T_{2E}^{ge}$ & Transmon Hahn-echo coherence time & 205.8 $\,\mu\mathrm{s}$ \\
$\bar{n}_{\text{th}}^q$ & Transmon thermal population & $0.25\%$ \\
$T_1^f$ & Transmon $|f\rangle$-state lifetime & $80.1\,\mu\mathrm{s}$ \\
$T_2^{gf}$ & Transmon $(|g\rangle + |f\rangle)$ coherence time & $45\,\mu\mathrm{s}$ \\
$g_a$ & Transmon–Alice vacuum Rabi coupling & $2\pi \times 5.841\,\mathrm{MHz}$ \\
$g_b$ & Transmon–Bob vacuum Rabi coupling & $2\pi \times 8.114\,\mathrm{MHz}$ \\
$\chi_e^a$ & Transmon–Alice dispersive shift & $-2\pi \times 71\,\mathrm{kHz}$ \\
$\chi_e^b$ & Transmon–Bob dispersive shift & $-2\pi \times 96\,\mathrm{kHz}$ \\
$\omega_r$ & Readout resonator frequency & $2\pi \times 8.379\,\mathrm{GHz}$ \\
$T_1^R$ & Readout resonator lifetime & $266\,\mathrm{ns}$ \\
$\chi_{e}^{r}$ & Transmon–readout dispersive shift & $-2\pi \times 411\,\mathrm{kHz}$ \\
$\omega_p$ & Purcell resonator frequency & $2\pi \times 8.529\,\mathrm{GHz}$ \\
$E_\text{J}$ & Josephson junction energy & $2\pi \times 24.46\,\mathrm{GHz}$ \\
$\chi_r^a$ & Readout-Alice cross-Kerr & $-2\pi \times 34\,\mathrm{Hz}$ \\
$\chi_r^b$ & Readout-Bob cross-Kerr & $-2\pi \times 131\,\mathrm{Hz}$ \\
$K_a$ & Alice mode self-Kerr & $-2\pi \times 2\,\mathrm{Hz}$ \\
$K_b$ & Bob mode self-Kerr & $-2\pi \times 19\,\mathrm{Hz}$ \\
\bottomrule
\end{tabular}
\end{adjustbox}
\caption{\textbf{System parameters and their values measured or inferred from the experiments.}}
\label{table:calibration_data}
\end{table}
\rowcolors{0}{}{}

\section{Cavity fabrication and loss characterization}
\label{app:bare_cavity}
The cavity is CNC-machined from high residual-resistivity-ratio (RRR$\sim300$) grade high-purity niobium in three separate parts, which are then electron-beam welded along the equators to eliminate seams at the joints. The cavity undergoes a buffered chemical polishing (BCP) process, removing approximately $120\,\mu$m of material to eliminate mechanical damage and embedded contaminants introduced during manufacturing. Afterward, it is cleaned with a high-pressure water rinse and vacuum-baked for three hours at $800^\circ \mathrm{C}$ to remove hydrogen absorbed into the niobium during the BCP treatment.

\begin{figure}
    \centering
    \includegraphics[width=\columnwidth]{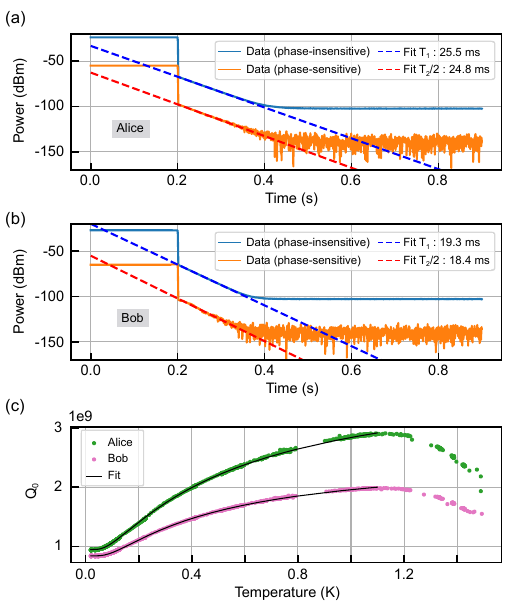}
    \caption{\textbf{Bare cavity characterization.}
    \textbf{(a)} Ringdown measurements for the Alice mode showing raw data and linear fits. 
    \textbf{(b)} Similar measurements for the Bob mode. 
    \textbf{(c)} Internal quality factors as a function of the mixing chamber plate's temperature for both modes and fitting to the TLS model.}
    \label{fig:bareT1}
\end{figure}

To assess the decay and dephasing time scales of the bare cavity modes in the milli-Kelvin regime, we perform ringdown measurements, applying a 200\,ms pulse and recording the reflected signal with a vector network analyzer. As depicted in Fig.~\ref{fig:bareT1}(a-b), we measure the emitted field from the cavity modes in two distinct ways--phase-insensitive (PI) and phase-sensitive (PS), using the methodology introduced in~\cite{Read2023}. The phase-insensitive trace is obtained by measuring the power (equivalent to the square sum of the cavity's I and Q quadratures) at each time point, then averaging over many such measurements. The resultant decay trace is well described by an exponential decay function characterized by the cavity's decay rate, 
\begin{align}
P_\text{PI} \propto\langle I^2 + Q^2 \rangle &\propto e^{-\frac{t}{T_1}}. 
\label{eq:phase_insensitive_avg}
\end{align}
From fitting to the PI power traces, we obtain the energy relaxation times of $25.545\,\text{ms} \pm 9\,\mu\text{s}$ for the Alice mode and $19.299\,\text{ms} \pm 4\,\mu\text{s}$ for the Bob mode.
The phase‑sensitive trace, on the other hand, requires the measurement of both amplitude and phase information of the cavity field. By ensemble averaging the $I$ and $Q$ quadratures separately, and then taking their squared sum, we obtain the PS trace that can be well-fitted by the following function, 
\begin{align}
P_\text{PS}\propto\langle I \rangle^2 + \langle Q \rangle^2 
&\propto e^{-\frac{t}{T_1}} \lvert\langle e^{i\theta(t)}\rangle \rvert^2\approx e^{-\frac{t}{T_2/2}}.
\label{eq:phase_sensitive_avg}
\end{align}  
Here, $\theta(t)$ is the random phase noise responsible for cavity dephasing. The last approximation in Eq.~\ref{eq:phase_sensitive_avg} assumes frequency-independent spectral density for $\theta(t)$ over the relevant bandwidth. Under this assumption, the overall decay of $P_\text{PS}$ reduces to a single exponential function, with a time constant of 
\begin{align}
\frac{T_2}{2} = \left(T_1^{-1}+2T_\phi^{-1}\right)^{-1},
\label{eq:T2_ringdown}
\end{align}  
where $T_\phi$ is the intrinsic dephasing time of the cavity. Therefore, the measurement of $T_1$ and $T_2$ times allows for the extraction of the pure dephasing times, which we find to be $(1700 \pm 181)\,\text{ms}$ and $(807 \pm 148)\,\text{ms}$ for the bare cavity modes. In these fittings, we only take the first 100\,ms of the trace to minimize the influence of the noise floor.

Additionally, we perform a temperature sweep to investigate the temperature dependence of the internal quality factor ($Q_0(T)$). This allows for the extraction of the zero-temperature loss tangent ($\delta_0$) associated with two-level systems (TLS), the residual surface resistance ($R_{\text{res}}$), and the temperature measurement efficiency ($\beta$) using the following equation~\cite{Romanenko2020},
\begin{equation}
\frac{1}{Q_0(T)}=F\delta_0 \tanh\Big(\beta \frac{\hbar \omega}{2k_BT}\Big)+\frac{R_{\text{res}}}{G}.
\label{eq:Q_vs_T}
\end{equation}
Here, $F$ is the surface oxide filling factor, and $G$ is the geometry factor,
\begin{equation}
F = \frac{\int_{\text{ox}} \epsilon_{\text{ox}}|\vec{E}|^2dv}{\int_{\text{all}} \epsilon_{\text{0}}|\vec{E}|^2dv},
\label{eq:F}
\end{equation}

\begin{equation}
G=\omega \mu_0 \frac{\int_{\text{all}}|\vec{H}|^2dv}{\int_{\text{sur}}|\vec{H}|^2d\sigma}.
\label{eq:G}
\end{equation}
The TESLA cavity geometry offers a notably low surface oxide filling factor ($F<2\times 10^{-8}$) and a large geometry factor ($G\approx 270~\Omega$), beneficial for enhancing cavity photon lifetime. At each temperature point, we perform a ring-down measurement to extract the loaded quality factor, $Q_L$. We then calculate the internal quality factor $Q_0$, from this $Q_L$ in conjunction with the coupling quality factor $Q_{\text{ext}}$ obtained from circle fitting at base temperature. We then fit the $Q_0$ vs.\ temperature data to Eq.~\ref{eq:Q_vs_T} and extract $F\delta_0$, $\beta$, and $R_{\mathrm{res}}/G$ from the fit. The factors $F$ and $G$ are obtained from finite-element method simulations using Eq.~\ref{eq:F} and Eq.~\ref{eq:G}, which allows us to determine $\delta_0$ and $R_{\mathrm{res}}$. The relevant parameters are shown in Table~\ref{tab:temp_sweep}. The measured residual surface resistance is higher than that typically observed in standard SRF cavities~\cite{Posen2020, Romanenko2017, Romanenko2020}, suggesting the presence of residual contaminants on the cavity surface. Subsequent studies with different cavities confirmed the likely presence of such contamination in the furnace used for the cavity preparation. Based on this, elimination of these additional losses can be achieved by an extra buffered chemical polishing (BCP) etching to remove the embedded contaminants. Furthermore, we plan to reduce the TLS losses coming from surface oxide via enhanced surface treatment techniques developed previously~\cite{Romanenko2020, Posen2020}. The most recent progress of SQMS (manuscript in preparation) includes demonstrating the ability to grow an air-exposure robust oxide with the drastically suppressed TLS density, which should allow a further multifold increase in the cavity quality factor. These combined improvements should allow at least an order of magnitude improvement in the energy relaxation times of both Alice and Bob modes.

\begin{table}[t]
\centering
\begin{tabular}{@{}cccccc@{}}
\toprule
$f_0$ (GHz) & $G$ $(\Omega)$ & $R_{\text{res}}(n\Omega)$ & $F \delta_0$ & $F$ & $\delta_0$ \\
\midrule
5.779 (Alice) & 295 & 73.2 & $8.0 \times 10^{-10}$ & $6.7 \times 10^{-9}$ & 0.12 \\
6.872 (Bob) & 298 & 116.5 & $7.9 \times 10^{-10}$ & $1.5 \times 10^{-8}$ & 0.05 \\
\bottomrule
\end{tabular}
\caption{\textbf{Extracted parameters from fitting the TLS model.}}
\label{tab:temp_sweep}
\end{table}
\section{Analysis of transmon-induced cavity losses}
\label{app:minimize_transmon_loss}

To preserve the excellent coherence properties of the cavity modes, we carefully minimize transmon chip-induced losses when introducing the transmon as the ancillary control element. In our device, the transmon chip is inserted through a narrow tunnel at the end of the 2-cell cavity, a feature reminiscent of the ``beampipe" found in accelerator-based TESLA cavities, where it serves as the passage for the particle beam. In our quantum device, however, this tunnel is engineered with a reduced diameter, forming a cylindrical waveguide with a high cutoff frequency well above the cavity’s fundamental modes. This design enables selective coupling between the cavity and the ancilla qubit, while effectively isolating the cavity field from other potential loss channels, such as radiation losses and seam losses that can arise with larger or open interfaces. To further support fast and high-fidelity qubit readout without compromising coherence, we incorporate an on-chip Purcell filter. This filter allows for a readout resonator linewidth of approximately $\kappa_r \approx 0.5$\,MHz, while maintaining readout-limited linewidths of $\kappa_q \approx 20$\,Hz for the qubit and $\kappa_{a,b} \approx 0.02$\,Hz for the cavity modes.

Following the methodology outlined in \cite{Ganjam2024}, we use finite-element simulations to calculate the loss channel participation ratios $p_{\text{bulk}}$, $p_{\text{MA}}$, $p_{\text{SA}}$, and $p_{\text{MS}}$ for the cavity modes,
\begin{equation}
p_{\text{bulk}} = \frac{\int_{\text{bulk}}\epsilon|\vec{E}|^2dv}{\int_{\text{all}}\epsilon|\vec{E}|^2dv},
\end{equation}
\begin{equation}
p_{\text{MA}}=\frac{t_{\text{surf}}\int_{\text{MA}}\epsilon_0|\vec{E}|^2d\sigma}{\epsilon_{\text{r,MA}}\int_{\text{all}}\epsilon|\vec{E}|^2dv},
\end{equation}
\begin{equation}
p_{\text{SA,MS}}=\frac{t_\text{surf}\int_{\text{SA,MS}}\epsilon|\vec{E}|^2d\sigma}{\int_{\text{all}}\epsilon|\vec{E}|^2dv}.
\end{equation}
Here, $\epsilon$ denotes the dielectric constant, $\epsilon_0$ is the vacuum dielectric constant, and $\epsilon_r$ is the relative dielectric constant, which we assume to be 10. Additionally, $t_{\mathrm{surf}}$ represents the thickness of the surface region, assumed to be 3~nm in this work.

Using these participation ratios and loss tangents, we estimate the cavity loss rate as $\kappa_{\text{bulk}} = \omega_c p_{\text{bulk}}\tan\delta_{\text{bulk}}$ from the bulk loss, and $\kappa_{\text{surf}}=\omega_c\sum_{i=\text{MS,MA,SA}}p_i\tan\delta_i$ from surface losses. We evaluate these loss rates as functions of the dispersive shift and transmon frequency, as summarized in Table~\ref{tab:loss_rate}. We target each dispersive shift for Alice and Bob to be within 50–100\,kHz to minimize transmon chip-induced losses while maintaining sufficient coupling for fast sideband control. Our analysis indicates that, under the current parameter configuration, transmon chip-induced material losses do not significantly limit cavity $T_1$ performance. 

\begin{table}[t]
\centering
\begin{adjustbox}{max width=\columnwidth}
\begin{tabular}{c|ccc|ccc}
\toprule
\multicolumn{1}{c|}{\textbf{Transmon}} 
& \multicolumn{3}{c|}{\textbf{Alice}} 
& \multicolumn{3}{c}{\textbf{Bob}} \\
\begin{tabular}[c]{@{}c@{}}$\omega_{ge}/2\pi$ \\ (MHz)\end{tabular}
& \begin{tabular}[c]{@{}c@{}}$\chi_e^a/2\pi$ \\ (kHz)\end{tabular} 
& \begin{tabular}[c]{@{}c@{}}$\kappa_{\text{bulk}}/2\pi$ \\ (Hz)\end{tabular} 
& \begin{tabular}[c]{@{}c@{}}$\kappa_{\text{surf}}/2\pi$ \\ (Hz)\end{tabular} 
& \begin{tabular}[c]{@{}c@{}}$\chi_e^b/2\pi$ \\ (kHz)\end{tabular} 
& \begin{tabular}[c]{@{}c@{}}$\kappa_{\text{bulk}}/2\pi$ \\ (Hz)\end{tabular} 
& \begin{tabular}[c]{@{}c@{}}$\kappa_{\text{surf}}/2\pi$ \\ (Hz)\end{tabular} \\
\midrule
6714 & -21  & 0.017 & 0.004 & -150 & 0.604 & 0.178 \\
6434 & -58  & 0.027 & 0.007 & -68  & 0.101 & 0.014 \\
6304 & -120 & 0.036 & 0.009 & -43  & 0.065 & 0.017 \\
\bottomrule
\end{tabular}
\end{adjustbox}
\caption{Transmon and substrate-induced loss rates for cavity modes.}
\label{tab:loss_rate}
\end{table}

The accurate determination of the transmon mode's effect on the loss channel participation ratios of the cavity mode depends critically on properly accounting for the nonlinearity-induced corrections, especially when the Lamb shift is comparable to the transmon-cavity detuning. We employ the Energy Participation Ratio (EPR) approach~\cite{Minev2021} to obtain the Lamb-shifted transmon frequency, and then renormalize the linear inductance of the Josephson junction used in the RF participation ratio simulations accordingly. This way, we are able to capture the transmon-cavity detuning precisely, allowing for a reliable estimation of the cavity's inherited loss from the transmon and the substrate material.

\section{Sideband pulse calibration}
\label{sideband_cal}

\begin{figure}[t]
    \centering
    \includegraphics[width=\columnwidth]{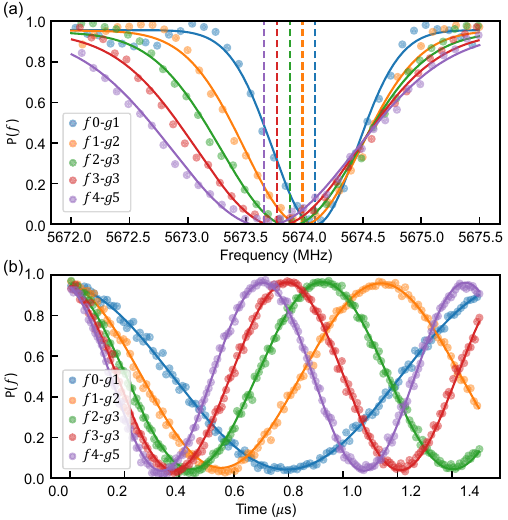}
    \caption{
    \textbf{Sideband pulse calibration}. \textbf{(a)} By initializing the state in $\lvert f,n\rangle$ and sweeping the sideband drive frequency, we observe the dip in the transmon $\ket{f}$ state population $P(f)$ that represents the corresponding sideband resonance frequency. \textbf{(b)} Driving the sideband transition at the measured resonance leads to the oscillations of $P(f)$, from which we extract the $\pi$-pulse time. We plot the first five sideband pulses for the Bob mode as an example, utilizing the SFP and PF techniques for enhancing the contrast of the calibration measurements.}
    \label{fig:sideband_cal}
\end{figure}

The sideband drive in our experiment is defined as a flat-top pulse, with 40~ns sine-squared ramp on each side to ensure adiabaticity. It is important to calibrate the resonance frequency as well as the $\pi$-pulse time for the sideband operation. Due to a drive amplitude-dependent Stark shift, the actual sideband transition frequency deviates from the bare resonance, \(2\omega_q + \alpha-\omega_c \). To precisely determine the transition frequency, we sweep the drive frequency around the bare resonance, and measure the resulting population in the $\ket{f}$ state. The exact resonance is then identified by a noticeable drop in the $\ket{f}$ population (Fig.~\ref{fig:sideband_cal}(a)). Once the precise transition frequency is identified, we measure time-domain Rabi oscillations on the transition $\ket{f,n} \leftrightarrow \ket{g,n+1}$ to calibrate the exact \(\pi\)-pulse duration (Fig.~\ref{fig:sideband_cal}(b)). The calibration process can be combined with the SFP and PF protocols, which enable calibration of the high-photon-number sideband transition frequencies and $\pi$ pulses without losing contrast. 

\section{Sideband-aided cavity reset operation}
\label{app:cavity_reset}
\begin{figure}[h]
    \centering
    \includegraphics[width=\columnwidth]{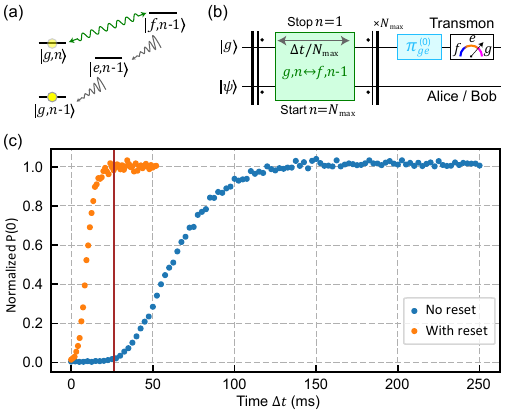}
    \caption{\textbf{Efficient cavity reset operation using the sideband interaction}. \textbf{(a)} The driven-dissipation process of the cavity population as a result of the cavity-transmon sideband interaction and the transmon decay. \textbf{(b)} Circuit diagram of the measurement for determining the duration of each sideband drive for the cavity reset operation. \textbf{(c)} Decay curves for a coherent state with $\alpha=4$ in the Alice mode, with (blue) and without (orange) using the sideband-aided reset protocol. The vertical line represents the duration of each sideband pulse for sufficiently resetting the cavity.}
    \label{fig:sideband_reset}
\end{figure}

We realize an efficient cavity reset operation in our experiment that requires a much shorter time than via the cavity's natural relaxation. As illustrated in Fig.~\ref{fig:sideband_reset}(a), this reset operation is activated by a driven dissipation process, where the cavity population hybridizes with the transmon via the cavity-transmon sideband interaction, thus inheriting the much faster decay rate from the transmon. In our experimental regime where the sideband interaction strength is much greater than the transmon's decay rate, the effective decay rate from $\lvert g,n\rangle$ to
$\lvert e,n-1\rangle$ is around half of the transmon $\lvert f\rangle$ to $\lvert e\rangle$ decay rate~\cite{Lu2017}. Thus, the engineered driven dissipation, through the cascaded decay process of $\lvert g,n\rangle \rightarrow
\lvert e,n-1\rangle \rightarrow \lvert g,n-1\rangle$, is happening at approximately the same rate as the transmon's decay, orders of magnitude faster than the cavity's relaxation. Using this protocol, we can reset an arbitrary initial cavity state to the vacuum state. Knowing the highest occupied Fock state $\lvert N_{\text{max}}\rangle$ in the cavity, we apply a sequence of sideband drives, going from $\ket{g,N_\text{max}}\leftrightarrow\ket{f, N_{\text{max}}-1}$ to $\ket{g,1}\leftrightarrow\ket{f,0}$, each lasting for a sufficiently long time compared with the transmon decay time. Given its autonomous nature, this protocol does not require knowing the exact initial state, and can tolerate small detunings of the sideband drives from resonance. 

Fig.~\ref{fig:sideband_reset}(b,\,c) demonstrates the circuit diagram and the result of the sideband-assisted reset protocol for the $\alpha=4$ coherent state in the Alice mode. In this case, we initiate the sideband drive at $N_\text{max} = 26$, as the cumulative population of all Fock states above $N_\text{max}$ is below 0.5\%. We apply the sequence of sideband drives for different cavity photon numbers, sweeping the duration of each pulse. We monitor the decay of the initial cavity state by measuring the cavity's vacuum state population, via a photon-number-selective $\pi$-pulse conditioned at the zero-photon peak at the end of the whole sequence. Compared to the natural decay curve that reaches an average photon number below 0.005 within approximately 160\,ms, the sideband-aided reset achieves the same photon number in only 25\,ms, corresponding to a duration time of 1\,ms for each sideband drive. This driven reset protocol can be further accelerated by parametrically coupling the cavity to the readout resonator with a large damping rate.

\begin{figure}[t]
    \centering
    \includegraphics[width=\columnwidth]{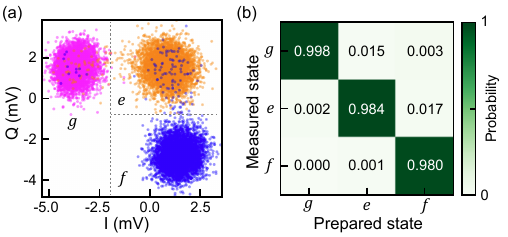}
    \caption{\textbf{Three-level readout of the transmon.} 
    \textbf{(a)} IQ blobs and demarcation lines for $\ket{g}, \ket{e}$ and $\ket{f}$ states. 
    \textbf{(b)} Resulting confusion (assignment) matrix.}
    \label{fig:cmatrix}
\end{figure}
\section{Transmon readout fidelity}
\label{app:rdt_gef}
The frequency, amplitude, and length of the readout pulse are optimized to maximize the 3-state discrimination probabilities for the transmon. A Josephson parametric amplifier (JPA) with about 17.7 dB gain is used to amplify the readout signal at the base temperature. The IQ blobs and demarcation lines for the $\ket{g}, \ket{e}$ and $\ket{f}$ levels are pictured in Fig.~\ref{fig:cmatrix}(a). This mapping is used for making decisions in SFP for correcting ancilla errors during Fock state preparations. The corresponding confusion matrix is shown in Fig.~\ref{fig:cmatrix}(b) resulting in an overall readout fidelity of approximately 98.8\%.

\section{Efficiency of the Sideband-Feedforward Protocol}
\label{app:SFP_efficiency}

\begin{figure}[t]
    \centering
    \includegraphics[width=\columnwidth]{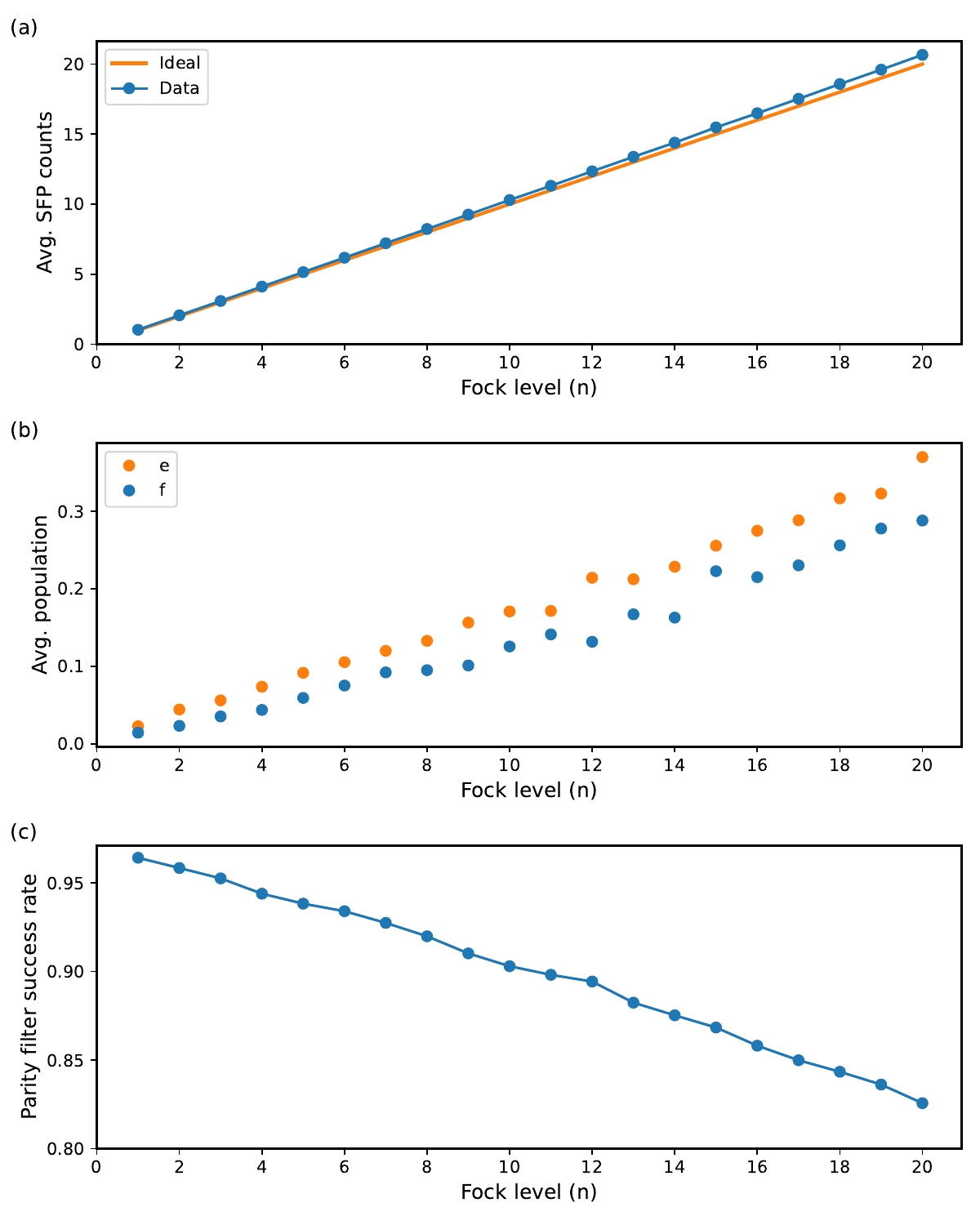}
    \caption{
    \textbf{Number of applied SFP operations for each Fock state.} \textbf{(a)} Average number of applied sideband-feedforward protocol (SFP) operations during the preparation of Fock states in Alice's mode. Ideally, the SFP is applied exactly once per sideband transition, assuming no errors occur (orange line). Experimentally, we find that the SFP is applied an average of $1.059$ times per desired sideband transition (blue line). \textbf{(b)} Measured average number of error syndrome states obtained during the SFP protocol for each targeted Fock state. \textbf{(c)} Measured success rate of the parity filter following the SFP operation. The filter rejects states with the incorrect parity, and the observed success rate is consistent with the Fock state preparation fidelity achieved using the SFP protocol alone.}
    \label{fig:SFP_efficiency}
\end{figure}
The high coherence time of the transmon ancilla, relative to the sideband operation time, ensures low error probability per sideband operation, allowing for the dramatic improvement in the final state fidelity without drastically increasing the time cost using the SFP. We characterize the efficiency of the SFP by measuring the number of feedforward operations necessary for each step up the Fock state ladder (see Fig.~\ref{fig:SFP_efficiency}). Experimentally, we apply SFP to prepare the cavity in the target Fock state, and measure how often additional feedforward operations are required. After repeating this measurement across many trials, we determine that feedforward is applied on average $1.059 \pm 0.003$ times per ladder step, verifying minimal overhead. Additionally, we find that 0.49\% of data shots cannot be corrected by the SFP, indicating errors outside the correctable subspace, such as photon decay events in the cavity. This low failure rate and minimal overhead confirm the high efficiency and robustness of our SFP implementation.

\section{Fidelity estimations for SFP}
\label{app:sfp_simulation}
Here we present an open-system numerical simulation of the Fock state preparation protocol discussed in Section III of the main text.
The simulation results indicate that the sideband feedforward protocol (SFP) nearly completely corrects decoherence errors in the ancilla transmon qubit. However, this comes at the cost of introducing additional errors due to transmon decoherence and measurement errors. This trade-off directly results in the dominant peak found in $|N-1\rangle$ of the out-of-target population, as shown in Fig.~2 and confirmed by simulation.
In the following subsections, we will explain the noise model used in the simulation and discuss the insights the simulation results provide.

\subsection{Noise model}
In the following, we will discuss a noise model for the transmon readout error and the decoherence effects of both the qudit and the ancilla transmon qubit based on the experimental parameters presented in Table \ref{table:calibration_data}. 

All unitary operations performed for the state preparation are rotations between two levels, $\hat{R}_{j, k} (\theta)$, where $j$ and $k$ are level indices of the qudit-transmon system and $\theta$ is the rotation angle. For example, the sideband transition is given as $\hat{R}_{(f, n), (g, n+1)}(\pi)$. 
Here, we construct the rotation unitaries as the qubit rotation around the $y$-axis between levels $j$ and $k$ while keeping other levels unchanged. This gives us
\begin{align}
\hat{R}_{j, k} (\theta) = & \,
\sum_{\ell \neq j, k} |\ell \rangle \langle \ell| + (|j\rangle \langle j| +|k\rangle \langle k|)\cos \frac{\theta}{2} \nonumber\\
& +(|k\rangle \langle j| - |j\rangle \langle k|) \sin \frac{\theta}{2} 
\end{align}
Alternative rotation unitaries can be constructed by, for example, taking the qubit rotation around the $x$-axis or assuming phase accumulation at other levels. Each alternative corresponds to a different pulse realization of the transition operation. However, the choice of construction does not affect the simulated fidelity of our Fock state preparation protocol. Therefore, we use the specific construction shown above for all rotations.
Without noise, the rotation operations modify the system state $|\psi \rangle $ such that
\begin{equation}
|\psi \rangle \rightarrow \hat{R}_{j, k}(\theta) |\psi \rangle.
\end{equation}

We use the Lindblad master equation to simulate the Fock state preparation protocol with noise.
The Lindblad master equation, which governs the time evolution of the system density matrix $\rho(t)$, can be written as 
\begin{equation}
\frac{\mathrm{d} \rho}{\mathrm{d} t} = -i [\hat{H}(t), \rho] + \sum_{\alpha} \mathcal{\hat{D}}[\hat{A}_\alpha] \rho.
\end{equation}
Here, $\mathcal{\hat{D}}[\hat{A}_\alpha]$ is the Lindblad superoperator given by $\mathcal{\hat{D}}[\hat{A}] \rho = \hat{A} \rho \hat{A}^{\dagger} - \tfrac{1}{2} \hat{A}^{\dagger} \hat{A}  \rho- \tfrac{1}{2}  \rho \hat{A}^{\dagger} \hat{A} $, each jump operator $\hat{A}_\alpha$ is associated with a decoherence channel to be discussed in the next paragraph, and $\hat{H}(t)$ is the system and control Hamiltonian generating the rotation operation such that
\begin{equation}
\hat{R}_{j, k} (\theta) = \mathcal{T} \exp{\left[-\frac{i}{\hbar} \int^{T}_0 \mathrm{d}t\ \hat{H}(t)\right]},
\end{equation}
where $T$ is the gate time and $\mathcal{T}$ is the time-ordering operator.
The noisy rotation operations $\mathcal{\hat{N}}_{j,k}(\theta)$ evolve the system density matrix such that
\begin{equation}
\rho \rightarrow \mathcal{\hat{N}}_{j,k}(\theta) \rho.
\end{equation}
By assuming that the rotation angle is directly proportional to the gate time, $\mathcal{\hat{N}}_{j,k}(\theta)$ can be approximated with a second-order symmetric trotterization \cite{Suzuki1991} such that
\begin{align}
\mathcal{\hat{N}}_{j,k} (\theta) 
\approx \Big( &
\mathcal{\hat{R}}_{j, k} \big(\tfrac{\theta}{2m}\big)
\Big[ 1 + \tfrac{T}{m} \textstyle\sum_\alpha \mathcal{\hat{D}}\big(\hat{A}_\alpha\big) \Big]
\nonumber\\
& \mathcal{\hat{R}}_{j, k} \big(\tfrac{\theta}{2m}\big) \Big)^m,
\end{align}
where $m$ is the trotterization step and $\mathcal{\hat{R}}_{j, k}(\theta)$ is the superoperator corresponding to the unitary rotation operation such that
\begin{equation}
\mathcal{\hat{R}}_{j, k}(\theta) \rho = \hat{R}_{j, k}(\theta) \rho \hat{R}_{j, k}^{\dagger}(\theta).
\end{equation}
We determined numerically that a value of \( m = 4 \) is adequate for our simulation.

Transmon depolarization and thermal heating for the ancilla transmon qubit with three levels are modeled by jump operators
\begin{align}
\hat{A}_{T_1^{q}, \downarrow} = & \sqrt{\gamma_{T_1^{ge}} (1 + \bar{n}_{\mathrm{th}}^q) }  (\sqrt{2} |e\rangle\langle f| + |g\rangle\langle e|), \\
\hat{A}_{T_1^{q}, \uparrow} = & \sqrt{\gamma_{T_1^{ge}} \bar{n}_{\mathrm{th}}^q } (\sqrt{2} |f\rangle\langle e| + |e\rangle\langle g|),
\end{align}
where $\gamma_{T_1^{ge}} = 1 / T_1^{ge}$ is the transmon decay rate, and $\bar{n}_{\mathrm{th}}^q$ is the average thermal population.
For transmon pure dephasing, we use the three-level system model presented in \cite{Li2012}, in which pure dephasing is described by two jump operators
\begin{align}
\hat{A}_{T_\varphi^{q}, f} & = \sqrt{2 \gamma_{T_\varphi^{ge}}} |f\rangle\langle f| , \\
\hat{A}_{T_\varphi^{q}, e} & = \sqrt{\gamma_{T_\varphi^{ge}}} |e \rangle\langle e|,
\end{align}
where $\gamma_{T_\varphi} = 1 / T_2^{ge} - 1 / (2 T_1^{ge})$ is the transmon pure dephasing rate.
The factor $\sqrt{2}$  associated with the $f$ state comes as an approximation from the ratio between the noise sensitivity of the $f-g$ energy difference and that of the $e-g$ energy difference, and more detailed discussions of the estimation of the dephasing time of the qubit can be found in \cite{Koch2007,Groszkowski2018}.
For the Alice-mode qudit, we write the jump operators for depolarization and pure dephasing as
\begin{align}
\hat{A}_{T_1^A} & = \sqrt{\gamma_{T_1^A}} \hat{a} , \\
\hat{A}_{T_\varphi^A} & = \sqrt{\gamma_{T_\varphi^A}} \hat{a}^{\dagger} \hat{a},
\end{align}
where $\gamma_{T_1^A} = 1 / T_1^A$ is the Alice-mode decay rate and $\gamma_{T_\varphi^A} = 1 / T_2^A - 1 / (2 T_1^A)$ is the Alice-mode pure dephasing rate.
The jump operators associated with the Bob-mode qudit can be written similarly.

We also take into account the transmon readout fidelity, which introduces error to the SFP, as the correction is conditional on the ancilla transmon qubit readout.
Although the confusion matrix shown in Fig.~\ref{fig:cmatrix} is sufficient for readout error mitigation by inverting the Markov process modeling noise \cite{Geller2020,Peters2023}, it does not distinguish between transmon-relaxation-induced readout error, which SFP can partially correct, and classification error, which SFP cannot correct.
To more accurately estimate the readout-related error, we model the readout fidelity using a transmon relaxation probability, denoted as \( p_\mathrm{readout} \), along with symmetric classification error probabilities. We use this model to simulate the experiment that measures the confusion matrix, and then optimize both the transmon relaxation probability and the classification error probabilities such that the simulated results best match the experimentally obtained confusion matrix. This yields the following results: \( p_\mathrm{readout}(e \rightarrow g) = 0.0055 \), \( p_\mathrm{readout}(f \rightarrow e) = 0.0110 \), and the classification error rate array
\begin{equation}
\begin{bmatrix}
0.9976 & 0.0024 & 0.0000\\
0.0024 & 0.9966 & 0.0010\\
0.0000 & 0.0010 & 0.9990
\end{bmatrix}.
\end{equation}
Note that the experimentally obtained confusion matrix used for optimization is slightly different from the one shown in Fig.~\ref{fig:cmatrix} due to parameter drift. In the simulation, we also take into account the qudit's decoherence during the transmon readout by including the corresponding depolarization and dephasing effects over the readout time of approximately $1.7\,\mu\text{s}$.

\subsection{Simulation results}
\begin{figure}
    \centering
    \includegraphics[width=\columnwidth]{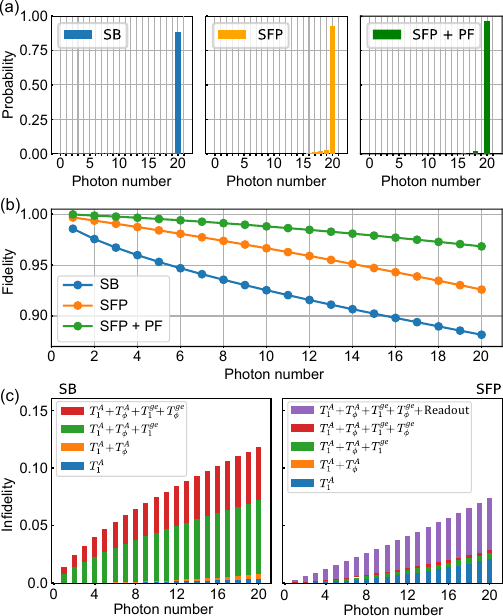}
    \caption{\textbf{Fock state preparation simulation for the Alice-mode qudit.} 
    \textbf{(a)} Simulation for preparing the $|n=20\rangle$ Fock state. 
    \textbf{(b)} Simulated fidelities improved by SFP and further improved by PF. 
    \textbf{(c)} Ceiling analysis of the error sources. 
} 
    \label{fig:app_sfp_simulation}
\end{figure} 
The simulated population distributions shown in Fig.~\ref{fig:app_sfp_simulation}(a) are consistent with the experimental results shown in Fig.~2(c). In particular, vanilla SB's result shows a nearly even distribution for the off-target states, while SFP's one exhibits a distinct peak at $|N-1\rangle$. This can be explained as follows: although the SFP effectively corrects most transmon errors, the lengthy readout time for each transmon (approximately 1700 ns) involved in the correction process makes uncorrected qudit decoherence errors more prominent. In contrast to the photon-number insensitive transmon errors, which result in nearly uniform off-target populations, the effective error rates for qudits tend to increase with the photon number. Consequently, the off-target distribution is skewed towards high photon number, with a peak at  $|N-1\rangle$.
In the simulation, the $|N-1\rangle$ peak is completely suppressed by PF, while it remains visible in the experimental data. This discrepancy arises because we do not account for the parity filtering efficiency in the simulation.

The predicted fidelities shown in Fig.~\ref{fig:app_sfp_simulation}(b) are higher than the experimental results presented in Fig.~2(d). This discrepancy is expected because our noise model is incomplete. In particular, control errors are not calibrated, and thus are not included in the noise model for simulation.
The absence of control errors in the noise model is likely the reason why the trend of the predicted SB's fidelities decreases at a slower rate than linear with increasing photon numbers. In contrast, the experimental fidelities show a faster-than-linear decrease as the photon number increases. In the simulation, the decoherence errors are approximately proportional to the sideband transition time, which decreases as $1 / |\chi_e| \sim 1 / \sqrt{N}$. However, as we move to higher photon numbers, we expect the control errors to become more pronounced as a result of the increasingly crowded energy levels. Therefore, it is likely that the control errors account for the faster-than-linear decrease in the experimental fidelities.

As we discussed earlier, SFP effectively corrects nearly all transmon decoherence errors, albeit at the cost of increasing qudit decoherence errors. This understanding is supported by the ceiling analysis illustrated in Figure \ref{fig:app_sfp_simulation}(c). In our ceiling analysis, we toggle specific noise channels on and off. For instance, the red bar demonstrates the difference in fidelities when transmon pure dephasing is on compared to when it is off. This approach allows us to estimate the contributions of transmon pure dephasing and other noise sources to the overall infidelities.
Comparing the SFP ceiling analysis on the right with the SB one on the left, the analysis clearly shows the effectiveness of SFP in correcting the transmon decoherence errors (red and green). Meanwhile, readout error (purple) and qudit depolarization (blue) become the limiting factors in the Fock state preparation protocol with SFP.
This indicates that achieving higher fidelities and preparing higher Fock states will require further improvements in the ancilla transmon readout fidelity and the qudit $T_1$, in addition to the control errors mentioned above.

\section{Raman-assisted beamsplitter interaction}
\label{app:Raman_assisted}
Here we derive the Alice-Bob beamsplitter interaction in the virtual Raman protocol. The Hamiltonian of the driven system is
\begin{align}
    \hat{\mathcal{H}}/\hbar = &\, \omega_a\hat{a}^\dagger \hat{a}+\omega_b\hat{b}^\dagger \hat{b}+\omega_{q0}\hat{q}^\dagger\hat{q}\nonumber\\ 
    &- E_J\cos_{\mathrm{nl}}\left[\theta_q\left(\hat{q}+\frac{g_a}{\Delta_a}\hat{a}+\frac{g_b}{\Delta_b}\hat{b}\right)+\mathrm{h.c.}\right]\nonumber \\ 
    &+ \sum_{k=1,2}\epsilon_k \cos\left(\omega_k t\right)\left(\hat{q}+\hat{q}^\dagger\right),
\end{align}
where $g_{a(b)}$ and $\Delta_{a(b)}$ stand for the coupling strength and detuning between Alice (Bob) and the transmon, and $\cos_\mathrm{nl}$ stands for fourth- and higher-order components of the cosine function. $\epsilon_{1(2)}$ and $\omega_{1(2)}$ are the amplitudes and frequencies of the two sideband drives acting on the transmon. Going into the displaced frame with respect to the sideband drives,
\begin{equation}
\begin{aligned}
    & \hat{\mathcal{H}}_\mathrm{disp}/\hbar = \omega_a\hat{a}^\dagger \hat{a}+\omega_b\hat{b}^\dagger \hat{b}+\omega_{q0}\hat{q}^\dagger\hat{q}
    \\
    & 
    \begin{aligned}
    - E_J\cos_{\mathrm{nl}} \Bigg[ 
    & \theta_q\Bigg( \hat{q}+\sum_{k=1,2}\xi_k e^{-i\omega_k t}+\frac{g_a}{\Delta_a}\hat{a}+\frac{g_b}{\Delta_b}\hat{b}\Bigg) \\
    & +\mathrm{h.c.} \Bigg],
    \end{aligned}
\end{aligned}
\end{equation}
where $\xi_{1,2}$ represents the drive-induced displacements. We further set the drive frequencies to be $\omega_{1(2)}=2\omega_{q0}+\alpha-\omega_{a(b)}-\Delta$ to match the VRBS resonance condition, where $\alpha$ is the transmon anharmonicity, and $\Delta$ is the virtual level detuning from $\lvert f00\rangle$. From going to the rotating frame and collecting the relevant slow-oscillating terms in $\cos_\mathrm{nl}$ expansion, we may approximate the virtual Raman (VR) Hamiltonian as
\begin{equation}
\begin{aligned}
    \hat{\mathcal{H}}_\mathrm{VR}/\hbar \approx \,
    & \frac{\alpha}{2}\hat{q}^\dag\hat{q}^\dag\hat{q}\hat{q} + \chi_a\hat{q}^\dag\hat{q}\hat{a}^\dag\hat{a} + \chi_b\hat{q}^\dag\hat{q}\hat{b}^\dag\hat{b}\\
    &-\frac{E_J\theta_q^4}{2} \hat{q}^{\dagger2}\left(\frac{\xi_1g_a}{\Delta_a}\hat{a}+\frac{\xi_2g_b}{\Delta_b}\hat{b}\right)e^{-i\Delta t}\\&-\frac{E_J\theta_q^4\xi_1\xi^*_2g_ag_b}{\Delta_a\Delta_b}\hat{a}\hat{b}^\dagger+\mathrm{h.c.},
\end{aligned}
    \label{eq:H_VR}
\end{equation}
where the third line represents the direct Alice-Bob beamsplitter interaction mediated by the 4th-order nonlinearity. Using second-order perturbation theory to eliminate the virtual states, we find the total effective beamsplitter interaction between Alice and Bob in the $\{\ket{10}, \ket{01}\}$ subspace to be 
\begin{align}
    \hat{\mathcal{H}}_\mathrm{VR}^{\mathrm{BS}}/\hbar \approx&  \left(\frac{E_J^2\theta_q^8\xi_1\xi^*_2g_ag_b}{2\Delta_a\Delta_b\Delta}-\frac{E_J\theta_q^4\xi_1\xi^*_2g_ag_b}{\Delta_a\Delta_b}\right)\hat{a}\hat{b}^\dagger \nonumber\\
    &+\mathrm{h.c.}\\
    \approx& \, 2\left(\frac{\alpha}{\Delta}+1\right)\frac{\alpha\xi_1\xi^*_2g_ag_b}{\Delta_a\Delta_b}\hat{a}\hat{b}^\dagger+\mathrm{h.c.}.
\end{align}
This clearly shows that, compared to the direct four-wave-mixing beamsplitter rate (represented by the 1 term), the VR scheme possesses a significant enhancement factor of $\alpha/\Delta$, making it over an order of magnitude faster than the former scheme in our experiment.

While the above derivation captures the essential interaction underlying the VR scheme, accurately modeling the system dynamics requires incorporating higher-order corrections induced by the drive or nonlinearity, which we demonstrate below.  We apply a Schrieffer-Wolff transformation~\cite{Shavitt1980} with a generator of $\hat{\mathcal{S}} = \sum_{n,m\geq0} (\hat{\mathcal{S}}_{n,m}^{(a)} + \hat{\mathcal{S}}_{n,m}^{(b)}$), where 
\begin{widetext}
\begin{align}
    \hat{\mathcal{S}}_{n,m}^{(a)} =&\tfrac{E_J\theta_q^4g_{a}\sqrt{6(n+1)}}{2\Delta_{a}[\Delta+\alpha+n\chi_h^{(a)} - (n+1)\chi_e^{(a)} +m\chi_h^{(b)} -m\chi_e^{(b)}]}|e,n+1,m\rangle \langle h,n,m | e^{i[\Delta+\alpha+n\chi_h^{(a)} - (n+1)\chi_e^{(a)} +m\chi_h^{(b)} -m\chi_e^{(b)} ]t} \nonumber\\
   & +\tfrac{E_J\theta_q^4g_{a}\sqrt{2(n+1)}}{2\Delta_{a}[\Delta+n\chi_f^{(a)}+m\chi_f^{(b)}]}|g,n+1,m\rangle \langle f,n,m | e^{i[\Delta+n\chi_f^{(a)}+m\chi_f^{(b)}]t} - \text{h.c.},
\end{align}
\begin{align}
    \hat{\mathcal{S}}_{n,m}^{(b)} =
    & \tfrac{E_J\theta_q^4g_{b}\sqrt{6(m+1)}}{2\Delta_{b}[\Delta+\alpha+n\chi_h^{(a)} - n\chi_e^{(a)} +m\chi_h^{(b)} -(m+1)\chi_e^{(b)}]}|e,n,m+1\rangle \langle h,n+1,m | e^{i[\Delta+\alpha+n\chi_h^{(a)} - n\chi_e^{(a)} +m\chi_h^{(b)} -(m+1)\chi_e^{(b)} ]t} \nonumber\\
   & + \tfrac{E_J\theta_q^4g_{b}\sqrt{2(m+1)}}{2\Delta_{b}[\Delta+n\chi_f^{(a)}+m\chi_f^{(b)}]}|g,n,m+1\rangle \langle f,n+1,m | e^{i[\Delta+n\chi_f^{(a)}+m\chi_f^{(b)}]t} - \text{h.c.}.
\end{align}
\end{widetext}
Here, $\chi^{(a,b)}_i$ denotes the dispersive shift of the Alice (a) or Bob (b) cavity mode conditioned on the ancilla being in state $|i\rangle$. Applying the Baker–Campbell–Hausdorff formula, the second line of Eq.~\eqref{eq:H_VR} transforms into
\begin{widetext}
\begin{equation*}
\begin{aligned}
&\frac{1}{2}\left[\hat{\mathcal{S}}, 
-\frac{E_J\theta_q^4}{2} \hat{q}^{\dagger2}\left(\frac{\xi_1g_a}{\Delta_a}\hat{a}+\frac{\xi_2g_b}{\Delta_b}\hat{b}\right)e^{-i\Delta t} + \text{h.c.}
\right] = \\
& \sum_{n,m} \beta_a \beta_b \sqrt{n{+}1} \sqrt{m{+}1}  |g\rangle\langle g| \otimes |n,m{+}1\rangle \langle n{+}1,m| 
\left(\tfrac{2}{\Delta + n \chi_f^{(a)} + m \chi_f^{(b)}} \right)\\
& - \beta_a \beta_b \sqrt{n{+}1} \sqrt{m{+}1}  |f\rangle\langle f| \otimes |n,m{+}1\rangle \langle n{+}1,m| e^{i(\chi_f^{(a)} - \chi_f^{(b)})t}
\left[ \tfrac{1}{\Delta + n \chi_f^{(a)} + (m{+}1) \chi_f^{(b)}} 
+ \tfrac{1}{\Delta + (n{+}1) \chi_f^{(a)} + m \chi_f^{(b)}} \right]  
\\
& + 3 \beta_a \beta_b \sqrt{n{+}1} \sqrt{m{+}1}  |e\rangle\langle e| \otimes |n,m{+}1\rangle \langle n{+}1,m| e^{i(\chi_e^{(a)} - \chi_e^{(b)})t} 
\\
& \times 
\left[ \tfrac{1}{\Delta+ \alpha +  n \chi_h^{(a)} - (n{+}1) \chi_e^{(a)} + m \chi_h^{(b)} - m \chi_e^{(b)} } 
+ \tfrac{1}{\Delta + \alpha + n \chi_h^{(a)} - n \chi_e^{(a)} + m \chi_h^{(b)} - (m{+}1) \chi_e^{(b)} } 
\right]
\\
& -3 \beta_a \beta_b \sqrt{n{+}1} \sqrt{m{+}1}  |h\rangle\langle h| \otimes |n,m{+}1\rangle \langle n{+}1,m| e^{i(\chi_h^{(a)} - \chi_h^{(b)})t} 
\\
& \times 
\left[ \tfrac{1}{\Delta + \alpha+ (n{+}1)\chi_h^{(a)} - (n{+}1)\chi_e^{(a)} + (m{+}1)\chi_h^{(b)} - (m{+}1)\chi_e^{(b)}} 
+ \tfrac{1}{\Delta + \alpha +  (n{+}1)\chi_h^{(a)} - (n{+}1)\chi_e^{(a)} + m \chi_h^{(b)} - (m{+}1) \chi_e^{(b)} } 
\right]\\
& + \text{h.c.},
\end{aligned}
\end{equation*}
\end{widetext}
where we define the quantities $\beta_{a,b} = \frac{E_J\theta_q^4\xi_{1,2}g_{a,b}}{2\Delta_{a,b}}$. 
From the equation above, we note that the beamsplitter rate is highly dependent on the transmon state, with detunings in the energy denominator differing by the anharmonicity for transmon $\lvert g\rangle$ and $\lvert e\rangle$ states. 
In general, the dispersive shifts are different for the two cavities. When the transmon is in the ground state, the resonant interaction is 
\begin{align}\label{eq:vrbs_g}
   \sum_{n,m} &
     \frac{E_J^2\theta_q^8\operatorname{Re}\left(\xi_1\xi^*_2\right)g_ag_b}{\Delta_{a}\Delta_{b}}
    \left( 
    \frac{|g\rangle\langle g|}{\Delta+n\chi_f^{(a)}+m\chi_f^{(b)}} 
    \right)
    \nonumber\\
    & \otimes
    \sqrt{(n+1)(m+1)}
    (|n, m+1\rangle \langle n+1, m|
    \nonumber\\
    & + |n+1, m\rangle \langle n, m+1|). 
\end{align}

This expression can be further simplified in the regime where $\chi_f^{(a,b)} \ll \Delta$:
\begin{align}
    &\frac{E_J^2\theta_q^8\operatorname{Re}\left(\xi_1\xi^*_2\right)g_ag_b}{\Delta\Delta_{a}\Delta_{b}}
    |g\rangle \langle g| \nonumber\\
    &\otimes (\hat{a}\hat{b}^\dag + \hat{a}^\dag\hat{b})\left[1-\frac{1}{\Delta}(\hat{n}_a\chi_f^{(a)} + \hat{n}_b\chi_f^{(b)})\right].
\end{align}
Intuitively, the beamsplitter rate is dependent on the cavity photon number because the drive detuning to the virtual states (VRBS detuning) for higher-photon-number sectors is altered by their dispersive shifts. This nonlinear correction to the beamsplitter gate has applications in universal qudit control, as discussed in SI~\ref{app:qudit_gates}.

Figure~\ref{fig:app_bs_matrix} shows a representative example of the VRBS operation in matrix form, where the effects on higher Fock states are obtained by numerically solving the Schrödinger equation (left panel). The block-diagonal structure of the evolution operator can be understood analytically from Eq.~\eqref{eq:H_VR}, whose corresponding unitary evolution is shown in the right panel. The analytic result shows good agreement with the exact numerical simulation, with small deviations likely resulting from higher-order processes that are neglected in the derivation.

\section{Numerical simulation of VR entanglement}
\label{app:Raman_assisted_numerical}
\begin{figure*}
    \centering
    \includegraphics[width=\textwidth]{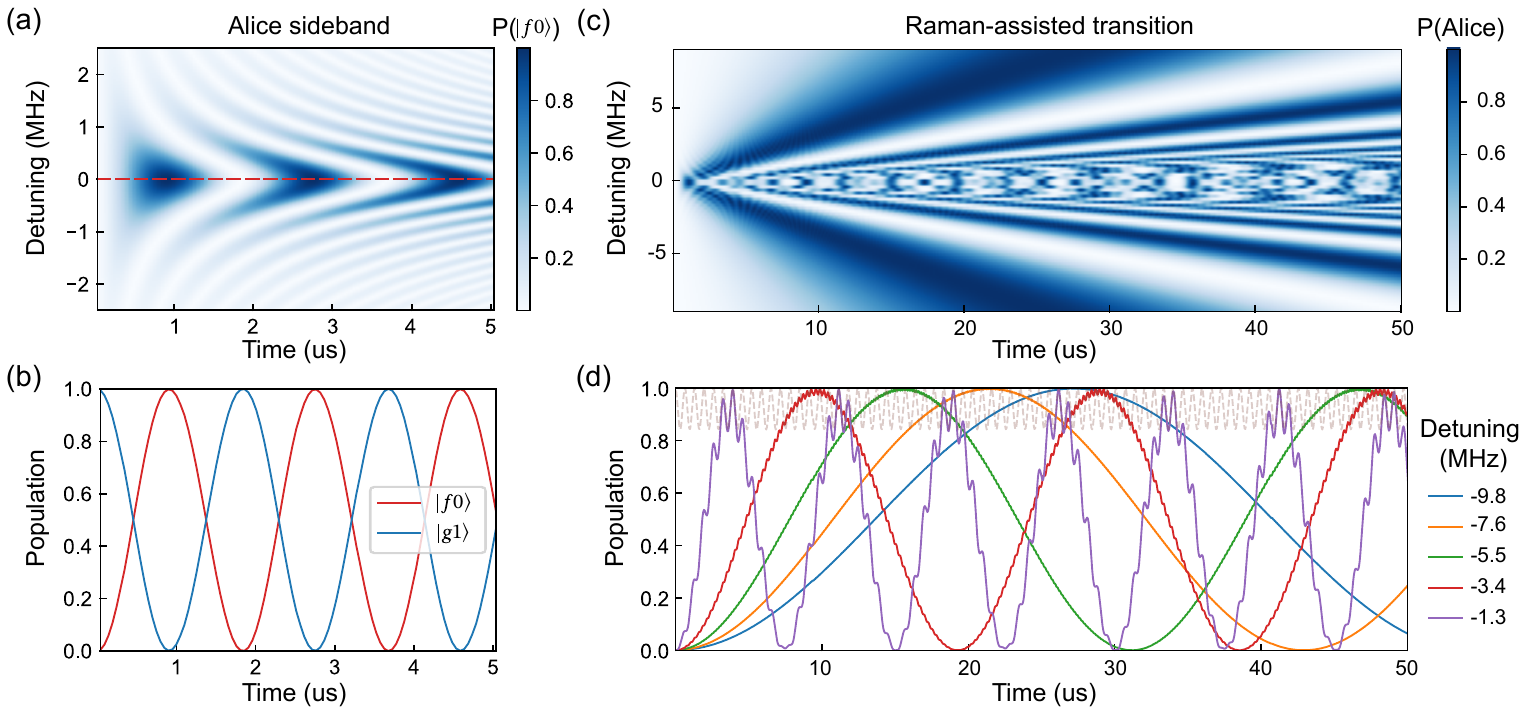}
    \caption{
    \textbf{Numerical simulations of Raman-assisted beamsplitter interactions.}
    \textbf{(a)} Simulated $\ket{f0} \leftrightarrow \ket{g1}$ sideband dynamics between Alice and the ancilla as a function of drive detuning with respect to the sideband resonance.
    \textbf{(b)} Line cuts taken along the resonant sideband condition indicated by the red dashed line in (a).
    \textbf{(c)} Simulated Raman-assisted population exchange between Alice and Bob, mediated by the ancilla and driven by dual sideband tones.
    \textbf{(d)} Line cuts of the cavity population dynamics at representative detunings. The dashed curve shows the transmon ground state population at a detuning of $-1.3$ MHz, which aligns with the modulation observed in the corresponding cavity population.}
    \label{fig:app_bs_simulation}
\end{figure*}

Here, we provide details on the numerical simulations of the Raman-assisted beamsplitter dynamics. The Hamiltonian in the bare basis is
\begin{equation}
    \begin{aligned}
    \hat{H}/\hbar =& \,\omega_a\hat{a}^\dagger \hat{a}+\omega_b\hat{b}^\dagger \hat{b}+\omega_{q}\hat{q}^\dagger\hat{q} \\
    & - E_J\cos_{\mathrm{nl}}\left[\theta_q \left(\hat{q}^\dag+\hat{q}\right)\right] \\ 
    & + g_a (\hat{a}^\dag+\hat{a}) (\hat{q}^\dag+\hat{q}) 
    + g_b (\hat{b}^\dag+\hat{b}) (\hat{q}^\dag+\hat{q}) \\
    &    + \sum_{k=1,2}\epsilon_k \cos\left(\omega_k t\right)\left(\hat{q}+\hat{q}^\dagger\right).
    \end{aligned}
\end{equation}
The circuit parameters are chosen such that the static energy spectrum of the Hamiltonian reproduces the experimental values listed in Table~\ref{table:calibration_data}.

Next, we calibrate the sideband drives for Alice and Bob individually. For a fixed sideband amplitude (e.g., $\epsilon_1$), we sweep the corresponding drive frequency $\omega_1$ to extract the resonant $\ket{f0}$–$\ket{g1}$ sideband oscillation period between Alice and the ancilla. This procedure is repeated until the simulated resonant oscillation period matches the experimentally observed value, thereby determining both the sideband drive amplitude $\epsilon_1$ and frequency $\omega_1$. As an example, Fig.~\ref{fig:app_bs_simulation}(a) shows the simulated sideband dynamics for Alice and ancilla. The red lines indicate the resonant conditions, and the corresponding frequency line cuts are shown in Fig.~\ref{fig:app_bs_simulation}(b).

With these calibrated parameters, we simulate the system evolution under simultaneous sideband drives, with identical detunings from their respective resonances. Figure~\ref{fig:app_bs_simulation}(c) displays the resulting population exchange between Alice and Bob, conditioned on the ancilla being in the ground state. For illustration, Fig.~\ref{fig:app_bs_simulation}(d) shows representative line cuts at several detunings. At large detunings, the population exhibits sinusoidal oscillations. As the detuning decreases, the oscillation frequency increases and develops a high-frequency modulation. This modulation originates from population in the transmon’s $|f\rangle$ state, due to the breakdown of the adiabatic condition at small sideband detuning. For instance, the transmon ground state population at a detuning of $-1.3$~MHz, shown as a dashed line, aligns with the modulation observed in the corresponding beamsplitter dynamics. Main-text Fig. 3(b) plots the extracted beamsplitter rates as a function of detuning (red line), showing excellent agreement with experimental data (blue dots).

While the above simulations capture the coherent dynamics, we also perform open-system simulations to study decoherence and its effect on gate fidelity. These are based on solving the Lindblad master equation with the following bare-basis Lindbladian:
\begin{equation}
    \begin{aligned}
        \hat{\mathcal{L}}[\rho] = & -\frac{i}{\hbar}[\hat{H}, \rho]
+ (T_1^{ge})^{-1} \hat{\mathcal{D}}[\hat{q}]\rho \\
& + \bar{n}^\text{q}_\text{th}(T_1^{ge})^{-1} \hat{\mathcal{D}}[\hat{q}^\dag]\rho 
 + 2(T_\phi^{ge})^{-1} \hat{\mathcal{D}}[\hat{q}^\dag\hat{q}]\rho  \\
   &+  (T_1^\text{A})^{-1} \hat{\mathcal{D}}[\hat{a}]\rho   +  (T_1^\text{B})^{-1} \hat{\mathcal{D}}[\hat{b}]\rho .
    \end{aligned}
\end{equation}
The decoherence parameters are set to: $T_1^{ge}=168\,\mu$s, $\bar{n}_\text{th}^\text{q}=4\%$, $T_{\phi}^{ge}=700\,\mu$s, $T_1^\text{A}=26$\,ms, and $T_1^\text{B}=20$\,ms, which are fitted to reproduce experimental results from individual sideband for Alice and Bob.
It is important to note that the ancilla thermal population and dephasing time used here differ from the values reported in Table~\ref{table:calibration_data}. This discrepancy arises because the Lindblad formalism assumes a white noise spectral density for ancilla pure dephasing, whereas real physical noise spectra are typically $1/f$ in the low-frequency regime. In driven systems, relevant decoherence rates are evaluated at different frequencies: for instance, dressed dephasing—which sets the ancilla’s thermal population and induces photon shot noise—samples the noise spectrum at the ancilla–drive detuning ($\sim$GHz); and sideband-dressed ancilla dephasing contributes to cavity decay at frequencies near the sideband detuning ($\sim$MHz). To mimic these effects within the white-noise Lindblad model, we manually adjust the ancilla thermal population to reflect the photon shot noise and reduce its pure dephasing rate to reflect the suppressed noise at larger frequencies.

\begin{figure}[t]
    \centering
    \includegraphics[width=\columnwidth]{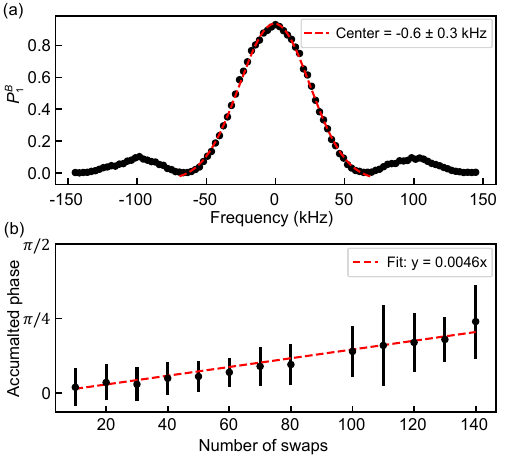}
    \caption{
    \textbf{VRBS drive calibration} \textbf{(a)} Bob's Fock $\ket{1}$ state population $P_1^B$ as a function of Bob sideband drive frequency with fixed duration, swept around the transition frequency $\omega_{\ket{f00} \leftrightarrow \ket{g01}} + \Delta$. The peak frequency corresponds to the difference in Stark shifts between the Alice and Bob modes. \textbf{(b)} Accumulated phase over multiple VRBS swap operations, measured by sweeping the phase of the final $U_{\text{BS}}(\phi)$ and identifying the phase correction that maximizes target state fidelity. 
    } 
    \label{fig:VRBS_Cal}
\end{figure}

To extract the cavity $T_1$ and $T_\phi$ from numerical simulations, we trace the ancilla to obtain the populations of Alice and Bob as a function of time. Then, we fit the decay oscillation with main-text Eq. (1).
Fig.~3(b) shows the extracted cavity $T_1$ and $T_\phi$ as functions of sideband detuning, which align well with the experiment. Both decay and dephasing times decrease as the system approaches the sideband resonance, consistent with enhanced decoherence due to the sideband drive. Notably, $T_\phi$ saturates at large detunings, a behavior attributed to photon shot noise arising from transmon-dressed dephasing. Because this dephasing mechanism samples the spectral density at the ancilla–drive detuning ($\sim$GHz), it is relatively insensitive to sideband detuning.

The simulated gate fidelity as a function of sideband detuning is displayed in Fig.~3(b) in the main text, showing a maximum near 5 MHz. The peak in the fidelity arises from a trade-off between gate speed and decoherence. Qualitatively, the beamsplitter time scales linearly with $\Delta$, while the sideband-dressed decoherence rate follows an approximate $\Delta^{-2}$ dependence. As a result, decreasing $\Delta$ enhances the interaction rate but more significantly increases decoherence, leading to an overall reduced fidelity. Conversely, at large detuning, the dominant decoherence mechanism shifts to photon shot noise arising from transmon-dressed dephasing, which is governed by the transmon–drive detuning ($\sim$GHz) and thus largely insensitive to MHz-scale variations in $\Delta$. In this regime, further increasing $\Delta$ leads to longer gate times without significant improvement in decoherence, again reducing the fidelity. The interplay of these two effects gives rise to the observed fidelity maximum, which is qualitatively reproduced by the numerical simulations.

\section{VRBS drives calibration}
\label{app:VRBS_calibration}
In this section, we discuss the calibration procedure for the VRBS drives. Accurate VRBS operation critically depends on precisely satisfying the frequency matching condition. Due to the drive-induced Stark shifts, the effective detuning of the VRBS drives differs from the detuning between Alice and Bob in the absence of driving. To account for this effect, we first prepare the Alice mode in the state $\ket{1}$. Subsequently, we simultaneously apply the $\ket{f00} \leftrightarrow \ket{g10}$ and $\ket{f00} \leftrightarrow \ket{g01}$ sideband drives with the same detuning at appropriate drive amplitudes. We keep the Alice sideband drive frequency fixed while sweeping the Bob sideband drive frequency. The precise resonance condition is then detected as the peak in the Bob $\ket{1}$ state population measurement (see Fig.~\ref{fig:VRBS_Cal}(a)).

The small but finite difference in Alice's and Bob's drive-induced Stark shifts results in a relative phase rotation between the Alice and Bob modes, accumulated over idling between concatenated VRBS pulses or ramping during the pulses~\cite{LuSchoelkopf2023}. For the error-detected VRBS swap operation, the transmon measurement after each swap can also contribute to a differential phase rotation of Alice and Bob due to their different cross-Kerr values with the readout mode (see Table~\ref{table:calibration_data}). Without proper calibration and compensation, these accumulated phases will translate to a coherent error that reduces the fidelity of the $(\lvert 01\rangle+\lvert 10\rangle)/\sqrt{2}$ state swap operation. In our experiment, we measure the phase rotation accumulated over a number of VRBS swap operations, by initializing the system in the $(\lvert 01\rangle+\lvert 10\rangle)/\sqrt{2}$ state, and varying the phase $\phi$ of the final VRBS beamsplitter operation that maps the entangled state back to the single-photon Fock state in Alice. The accumulated phase is identified by the angle that maximizes the target state fidelity. We observe a linear increase of the accumulated phase with respect to the number of swap operations, where we extract the phase shift per gate to be $0.0046$\,rad, see Fig.~\ref{fig:VRBS_Cal}(b). This phase $\phi$ is then applied to the beamsplitter operation $U_\text{BS}(\phi)$ for the error-detected swap operation of the state $(\lvert 01\rangle+\lvert 10\rangle)/\sqrt{2}$.

\section{Transmon heating detection after each swap operation} 
\label{app:transmon_heating}

\begin{figure}[t]
    \centering
    \includegraphics[width=\columnwidth]{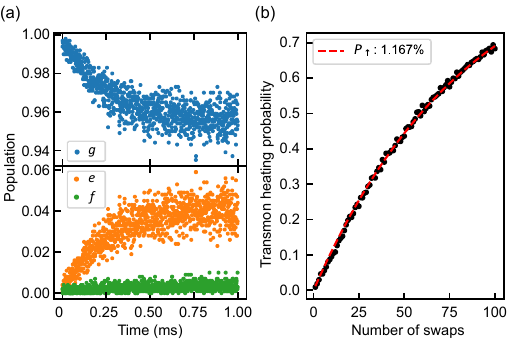}
    \caption{
    \textbf{Transmon heating during VRBS drive} \textbf{(a)} Transmon population during continuous VRBS drive. The transmon heats up, reaching a thermal population of $\sim4\%$  after a 1\,ms VRBS drive. \textbf{(b)} The probability of the transmon heating event up to 100 swap gate operations. We estimate the transmon heating probability as $1.167\pm0.003$ \% per swap gate operation.
    }
    \label{fig:transmon_heating}
\end{figure}

Transmon heating events during swap operation prevent coherent oscillation between Alice and Bob, thereby causing dephasing. To detect and characterize these heating events, we measure the transmon state immediately after each swap operation. We perform measurements up to 100 consecutive swap operations, post-selecting out any instances where heating is detected. We fit the transmon heating detection probability as a function of number of swaps, $P(n) = 1 - (1-P_\uparrow)^n$, to estimate the transmon heating rate per swap operation, $P_\uparrow$. Figure \ref{fig:transmon_heating} illustrates the heating detection rate observed over these 100 swap operations. From these measurements, we estimate a heating probability of approximately $P_\uparrow \approx 1.167\pm0.003\%$ per swap operation. This heating (and decay) of the driven ancilla results in stochastic hopping of the ancilla population between the ground and excited state, further leading to the dephasing of the VRBS process due to the large dispersion of the effective VRBS rate over different ancilla states (see SI~\ref{app:Raman_assisted}). Therefore, by adding mid-circuit transmon measurements, we are able to detect the dominant dephasing error in VRBS, and demonstrate an improvement in the VRBS fidelity by post-selecting on no heating events, see main text Fig.~3. 

\begin{figure}[h]
    \centering
    \includegraphics[width=\columnwidth]{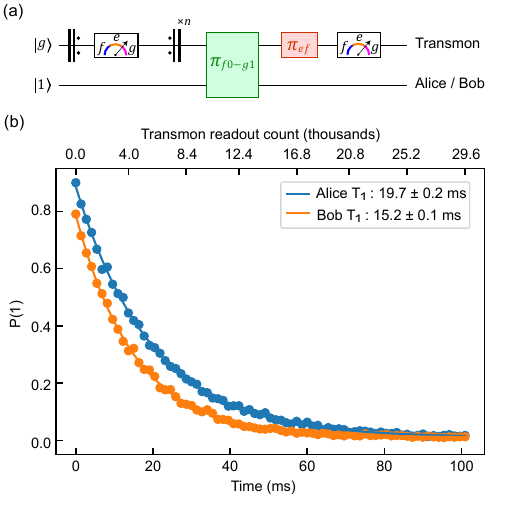}
    \caption{
    \textbf{Cavity $T_1$ as a function of transmon measurement counts.} \textbf{(a)} The quantum circuit showing the experimental protocol for measuring the cavity $T_1$ in the presence of multiple transmon readouts during the wait period. \textbf{(b)} The relaxation times are not affected by transmon readouts.
    }
    \label{fig:cav_T1_transmon_meas}
\end{figure}

To verify that repeated transmon measurements do not degrade the cavity lifetime, we follow the protocol described below. First, we prepare the cavity in the single‑photon Fock state $\ket{1}$. We then perform $N$ consecutive transmon readout operations, each lasting for $t_r = 3.4\,\mu$s with $N$ ranging from 0 to 29600. After the readout-delay sequence, we map the remaining Alice $\ket{1}$ population onto the transmon excited state via a sideband $\pi$-pulse and a transmon $\pi_{ef}$ pulse, followed by transmon population measurement. Converting the number of readout-delay cycles to an effective elapsed time $t = N(t_r + \Delta t)$, we extract the cavity population $P_{1}(t)$ over time, and fit it to an exponential decay, as shown in Fig.~\ref{fig:cav_T1_transmon_meas}. Within experimental uncertainty, the extracted decay time agrees with the cavity $T_{1}$ measured without interleaved transmon readouts, confirming that repeated transmon measurements do not contribute to additional photon loss in the cavity.

\section{Mapping dual-rail cavity states onto transmon states}
\label{app:cavity_mapping}

\begin{figure}[b]
    \centering
    \includegraphics[width=\columnwidth]{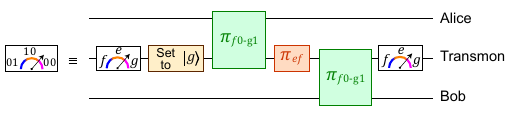}
    \caption{\textbf{Mapping of cavity states to the transmon.} The transmon is first reset to the $\ket{g}$ state by measurement and following conditional $\pi$ pulse(s). Then a combination of sideband and transmon pulses are applied to perform the mapping: $\ket{00} \rightarrow \ket{g}, \ \ket{10} \rightarrow \ket{e}$ and $\ket{01} \rightarrow \ket{f}$.}
    \label{fig:mapping}
\end{figure}

\begin{figure*}[]
    \centering
    \includegraphics[width=0.9\textwidth]{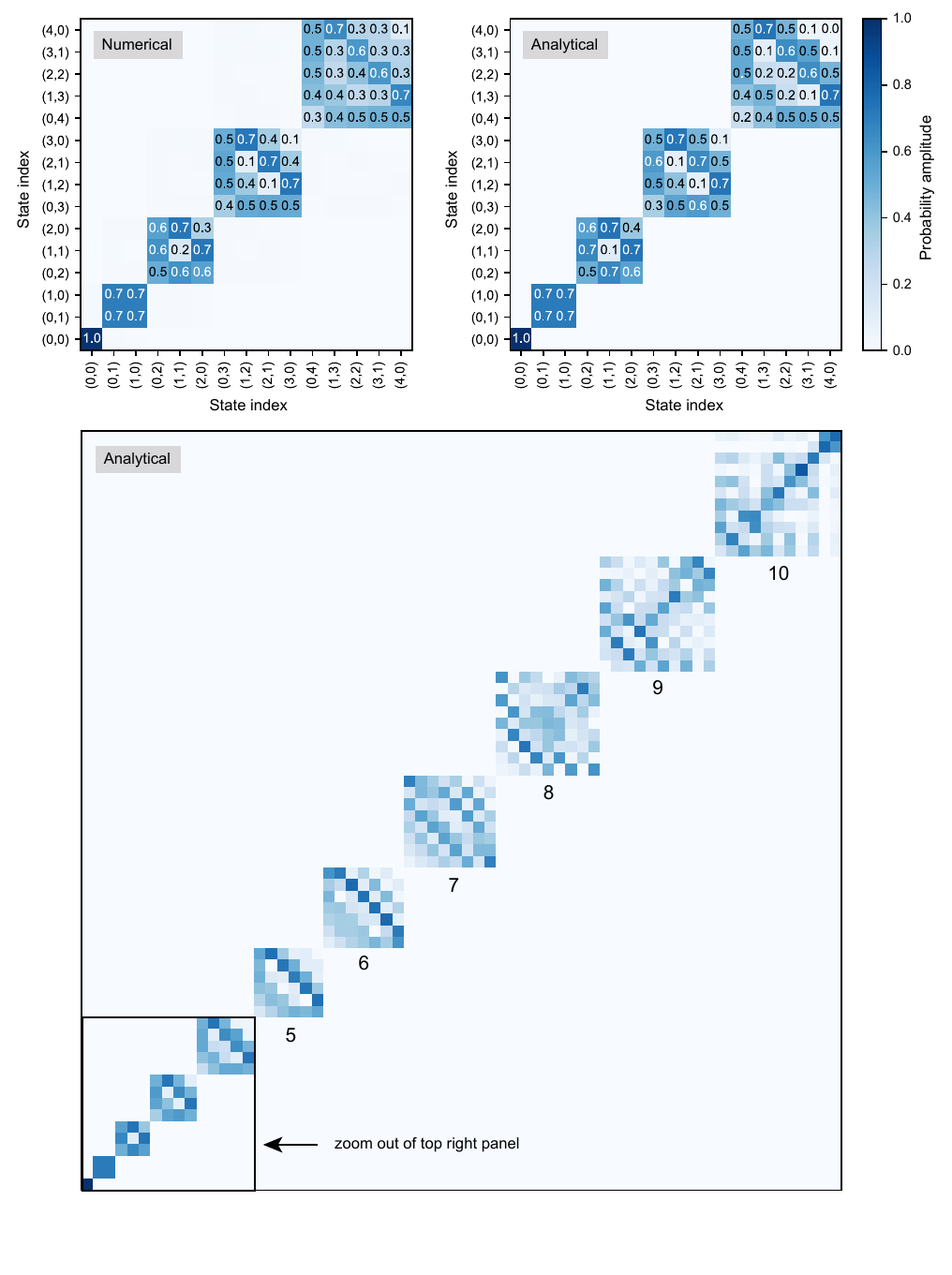}
    \caption{
    \textbf{Matrix elements of the VRBS operation.} 
    The matrix takes a block-diagonal form, reflecting the structure of the VRBS gate when acting on higher Fock states. The top left panel shows the exact numerical result obtained by solving the time-dependent Schrödinger equation, while the top right panel displays the corresponding analytical unitary evolution derived from Eq.~\eqref{eq:vrbs_g}. The two results show good agreement. 
    The bottom panel extends the analytical result to up to 10 photons. 
    The simulation is performed at a sideband detuning of $-5.5$ MHz. Only the probability amplitudes are shown for brevity.}
    \label{fig:app_bs_matrix}
\end{figure*}

To measure the populations of single-photon states in the Alice and Bob modes using single-shot readout, we map the cavity states onto the transmon states. The circuit diagram is illustrated in Fig.~\ref{fig:mapping}. After resetting the transmon to the ground state, we apply a sideband transition pulse $\ket{g10} \leftrightarrow \ket{f00}$ (corresponding to the Alice mode). Subsequently, we apply a photon-number-unselective $\pi_{ef}$ pulse with a short duration ($\Delta t \ll 2\pi/\chi_e^a,\,2\pi/\chi_e^b$). Finally, we apply a second sideband transition pulse $\ket{g01} \leftrightarrow \ket{f00}$ (corresponding to Bob mode). Through this sequence, the population in the $\ket{10}$ state maps to the transmon $\ket{e}$ state, the $\ket{01}$ state population maps to $\ket{f}$, and the $\ket{00}$ state maps to $\ket{g}$. This method enables single-shot measurement of $\ket{00},\ \ket{01},\ \ket{10}$ states within a duration significantly shorter ($\ll 1/\chi_e$) compared to parity measurement or PNRS, which typically have durations on the order of $1/\chi_e$. The resulting confusion matrix $M_{\rm map}$ for this mapping is 
\begin{equation}
\label{eq:map_confusion}
M_{\rm map} = 
\begin{bmatrix}
0.9959 & 0.0491 & 0.0100\\
0.0018 & 0.9467 & 0.0172\\
0.0022 & 0.0042 & 0.9728
\end{bmatrix},
\end{equation}
where the columns (rows) correspond to the prepared (measured) states of $\ket{00}, \ket{10}$ and $\ket{01}$. The observed asymmetry between Alice and Bob arises from the order of the mapping procedure. We use this confusion matrix to correct the cavity state readout errors for the beamsplitter operations.

\section{Qudit gates}
\label{app:qudit_gates}

\begin{figure*}[t]
    \centering
    \includegraphics[width=\textwidth]{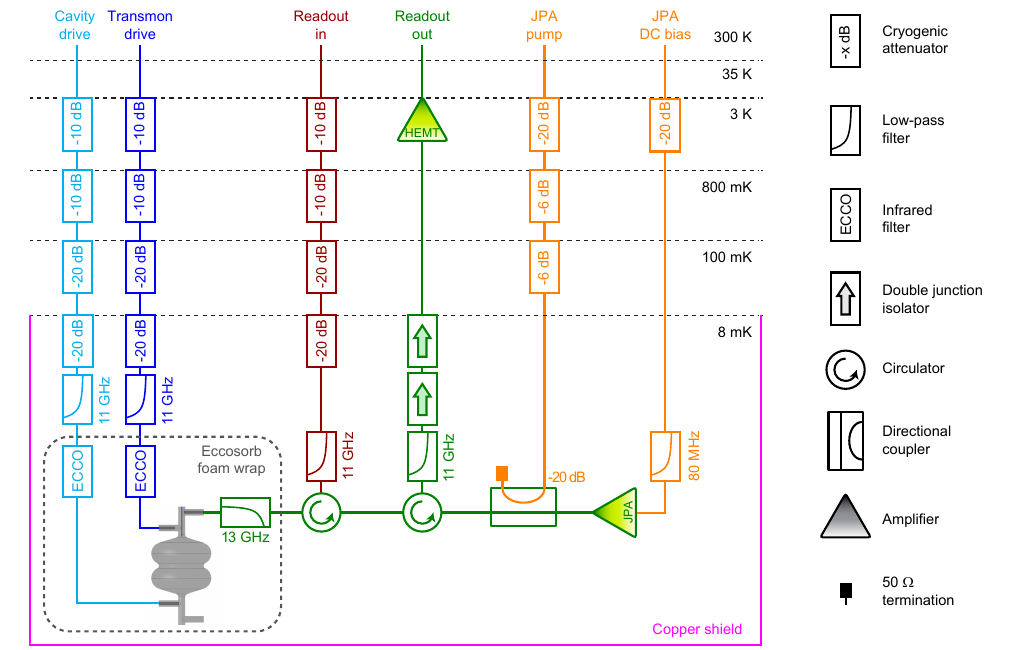}
    \caption{\textbf{Simplified cryogenic wiring diagram.}}
    \label{fig:wiring}
\end{figure*}

Universal control of a two-qudit system requires the ability to perform arbitrary single-qudit rotations along with at least one entangling operation~\cite{Braunstein2005CV}. In our multimode architecture, single-qudit rotations $R_d(\vec{\theta})$ can be implemented using either sideband gates~\cite{law1996arbitrary,Liu2021qudit} or echoed conditional displacement gates~\cite{You2024crosstalk}. The VRBS interaction discussed in the main text provides a mechanism for generating entanglement between higher Fock states as well. This operation takes a block-diagonal form, where each basis state $\ket{m,n}$ is rotated within a photon-number-conserving subspace—i.e., among states $\ket{m',n'}$ satisfying $m'+n' = m+n$. Together with single-qudit rotations, this operation enable universal control within a finite-dimensional Hilbert space constructed from $d$-dimensional qudits.  For example, universal operations in the two-qudit Hilbert space can be performed by:
\begin{align}
\label{eq:universal_qutrit}
    \mathcal{U}= & \prod_{j=1}^{N_{\rm block}}\left[\textsc{R}_d(\vec{\theta}_j) \otimes \textsc{R}_d(\vec{\phi}_j) \cdot \textsc{VRBS} \right] \nonumber \\
    &\ \cdot \textsc{R}_d(\vec{\theta}_0) \otimes \textsc{R}_d(\vec{\phi}_0),
\end{align}
where $N_{\rm block}$ is the number of blocks of those gates within the square brackets.

With the results of Fig.~\ref{fig:app_bs_matrix}, it is possible to investigate the properties of VRBS as a two-qutrit entangling gate, as a concrete example. 
For qutrits, it has a Schmidt rank of 9. The entangling power, $e_p(U)$ can be defined as the average of the linear entropy over product states $\ket{\Psi}=\ket{\psi}_1\otimes\ket{\psi_2}$ following~\cite{wang2003entangling}:
\begin{equation}
    E(\ket{\Psi})=1-\text{Tr}_1 \rho_1^2,
\end{equation}
where $\rho_1=\text{Tr}_2(U\ket{\Psi}\bra{\Psi}U^\dagger)$ is the reduced density matrix, and $U$ is the entangling gate being investigated. We estimate $e_p(\rm VRBS)=0.379(2)$ by sampling $10^5$ $\ket{\Psi}$, similar to the standard qutrit CSUM gate (${\rm CSUM_3}|m,n\rangle = \ket{m,m+n \ {\rm mod} \ 3 }$) which is known analytically to have $e_p(\text{CSUM}_3)=3/8$~\cite{wang2003entangling}. This implies that VRBS is an efficient qutrit entangling gate.  Further, by numerical optimization of the $SU(3)$ angles in Eq.~(\ref{eq:universal_qutrit}), approximate synthesis of the CSUM$_3$ gate from $2\leq N\leq 6$ blocks were identified with fidelities
\begin{equation}
    \mathcal{F}_{\text{CSUM}_3}^{N_{\rm blocks}}=\{51\%,88\%,92\%,99.1\%,99.999974\%\}.
\end{equation}
Demonstration of universality in our 2-cell system is the topic of future work, with emphasis placed on qudit CSUM gates.

\section{Experimental setup}

Figure~\ref{fig:wiring} illustrates the cryogenic wiring diagram of our measurement setup. The microwave components and design choices are carefully optimized to minimize thermal excitation of both the cavity modes and the transmon. The double-cell cavity is mounted vertically within a custom-built support structure—or cage—comprising straight and semicircular clamps made from oxygen-free high thermal conductivity (OFHC) copper, ensuring efficient thermal anchoring and heat extraction.

To suppress infrared (IR) radiation, the cage is wrapped in Eccosorb\textsuperscript{TM} foam, which absorbs stray IR photons that could otherwise be absorbed by the cavity or surrounding microwave components. Two commercial Eccosorb filters, thermalized via OFHC copper braids, are installed on the drive lines to the cavity and transmon, providing attenuation of high-energy IR photons coming down the coaxial cables. A High-Energy-Radiation-Drain (HERD-2) filter is placed on the readout line to block infrared radiation while minimizing signal attenuation, ensuring high-fidelity readout. All IR filters are mounted inside the Eccosorb-wrapped region for protection against ambient IR photons.

A copper can is mounted on the dilution refrigerator’s base plate to enclose and shield all components thermalized to the base temperature at around 8\,mK. To further reduce susceptibility to magnetic noise, a $\mu$-metal shield is placed around the vacuum chamber, protecting the system from external magnetic fields.

For control and measurement, we use the OPX+ and Octave RF hardware from Quantum Machines. The transmon is read out in reflection via its on-chip readout resonator. A Josephson Parametric Amplifier (JPA) is used as the first-stage phase-insensitive amplifier in the readout chain. The JPA’s charge pump (8.375\,GHz) and DC bias are supplied by a Rohde \& Schwarz SMA100B signal generator and a Yokogawa GS200 DC voltage/current source, respectively.

\bibliography{main}